\documentclass[reprint]{aastex631}

\accepted{for publication in The Astrophysical Journal}

\shorttitle{Polarization of Self-luminous Exoplanets}
\shortauthors{Chakrabarty, Sengupta \& Marley}
\graphicspath{{figures/}}
\usepackage{natbib}
\usepackage{amsmath}
\usepackage{wasysym}
\begin{document}
\renewcommand{\vec}[1]{\boldsymbol{#1}}
\newcommand{\mvec}[1]{$\boldsymbol{#1}$}
\renewcommand{\vector}[4]{\begin{bmatrix} #1\\ #2\\ #3\\ #4\end{bmatrix}}\renewcommand{\matrix}[6]{\begin{bmatrix} #1&#2&0&0\\#2&#3&0&0\\0&0&#4&#5\\0&0&$-$#5&#6 \end{bmatrix}}
\newcommand{\matrixz}[5]{\begin{bmatrix} #1&#2&0&0\\#2&#3&0&0\\0&0&#4&0\\0&0&0&#5 \end{bmatrix}}
\newcommand{\I}{$I$ }
\newcommand{\Q}{$Q$ }
\newcommand{\U}{$U$ }
\newcommand{\V}{$V$ }
\renewcommand{\deg}{$^\circ$ }
\renewcommand{\arccos}[1]{\cos^{-1}\left(#1\right)}
\newcommand{\muphi}{$(\mu,\phi) $}
\newcommand{\muphip}{$(\mu',\phi') $}
\newcommand{\muphiz}{$(-\mu_0,\phi_0) $}
\newcommand{\Mphi}{$(M,\Phi) $}

\title{Polarization of Rotationally Oblate Self-Luminous Exoplanets with Anisotropic Atmospheres}

\correspondingauthor{Aritra Chakrabarty}
\email{aritra@iiap.res.in}

\author[0000-0001-6703-0798]{Aritra Chakrabarty}
\affil{Indian Institute of Astrophysics, Koramangala 2nd Block,
Sarjapura Road, Bangalore 560034, India}
\affil{University of Calcutta, Salt Lake City, JD-2 Kolkata 750098, India}

\author[0000-0002-6176-3816]{Sujan Sengupta}
\affil{Indian Institute of Astrophysics, 
Koramangala 2nd Block, Sarjapura Road, Bangalore 560034, India}

\author[0000-0002-5251-2943]{Mark S. Marley}
\affil{Lunar and Planetary Laboratory, University of Arizona, Tucson, Arizona 85721, USA}

\begin{abstract}

Young self-luminous giant exoplanets are expected to be oblate in shape owing to the high rotational speeds observed for some objects. Similar to the case of brown dwarfs, the thermal emission from these planets should be polarized by scatterings of molecules and condensate cloud particles, and the rotation-induced asymmetry of the planet's disk would yield to net non-zero detectable polarization.  Considering an anisotropic atmosphere, we present here a three-dimensional approach to estimate the disk-averaged polarization that arises due to the oblateness of the planets.  We solve the multiple-scattering vector radiative transfer equations at each location on the planetary disk and calculate the local Stokes vectors and then calculate the disk-integrated flux and linear polarization. For a cloud-free atmosphere, the polarization signal is observable only in the visible wavelength region. However, the presence of clouds in the planetary atmospheres leads to a detectable amount of polarization in the infrared wavelength region where the planetary thermal emission peaks. Considering different broad-band filters of the SPHERE-IRDIS instrument of the Very Large Telescope, we present generic models for the polarization at different wavelength bands as a function of their rotation period. We also present polarization models for the Exoplanets $\beta$ Pic b and ROXs 42B b as two representative cases which can guide future observations. Our insights on the polarization of young giant planets presented here would be useful for the upcoming polarimetric observations of the directly imaged planets.

\end{abstract}

\keywords{planets and satellites: atmospheres --- radiative transfer --- polarization --- scattering --- infrared: planetary systems}

\section{Introduction} \label{sec:intro}

The polarimetric technique has been gaining momentum over the past few years in the field of exoplanet characterization, especially for the young directly imaged planets. The use of adaptive optics coronagraphic systems has enabled us to conduct direct photometric, spectroscopic as well as polarimetric observations of the substellar companions \citep[e.g.,][]{bryan18, miles19, jensen20}. Polarimetric observations of brown dwarfs and the directly imaged planets using highly sensitive instruments have already been reported by \cite{millar20, jensen20, holstein21}, among others. The increasing number of reports on polarization observations of sub-stellar mass objects calls for a better understanding of the atmospheric processes that give rise to polarization. The thermal radiation of the objects becomes linearly polarized due to scattering by atmospheric molecules and cloud particles. The detected net non-zero disk-integrated polarization, on the other hand,  is attributed to the asymmetry of the visible disk due to various reasons such as oblateness, inhomogeneous or patchy cloud coverage in the atmosphere, gravitational darkening, etc. \citep[e.g.,][]{sengupta05, sengupta09, sengupta10, marley11, dekok11, stolker17, sanghavi18}. Polarimetric observations of the red dwarf stars and brown dwarfs \citep{miles15, millar20}  strongly suggest that the observed polarimetric variations correlate with the rotation-induced oblateness of those objects. Hence, in this paper, we focus on the estimation of the polarization caused solely by the rotation-induced oblateness of the young self-luminous gas giant planets. 

Theoretical computation of the flux and polarization observable from a directly imaged Exoplanet plays a pivotal role in guiding future polarimetric missions. These forward models can be used to relate the observed flux and polarization from the planets to their different physical and atmospheric properties. The high-resolution spectroscopic studies \citep[e.g.,][]{snellen14, bryan18, xuan20} provide information about the line-of-sight (LOS) component of the equatorial rotation velocity ($v_e\sin i$) of the brown dwarfs and the directly imaged giant Exoplanets. However, the values of the inclination angles ($i$, not to be confused with the orbital inclination angle) of the rotation axis with respect to the observer cannot be found out from such studies to date. Polarimetric observations can complement such spectroscopic as well photometric observations and break the degeneracies among the estimated properties such as rotation speed ($v_e$), the inclination angle of the rotation axis ($i$), surface gravity ($g$), among others. 

The net non-zero disk averaged polarization of a substellar object can be attributed to its rotation-induced oblateness even for a cloud-free atmosphere, as in the case of T-dwarfs \citep[e.g.,][]{sengupta09}. However, polarization from a cloud-free atmosphere is predominantly caused by the Rayleigh scattering of the thermal emission by the atoms and molecules of the gases present in the atmosphere. Hence, such a polarization signal is detectable only in the visible wavelength region.  In the presence of condensate cloud or haze particulates in the atmospheres, a significant amount of polarization arises in the infra-red wavelength region where the brightness of the objects peaks \citep[e.g.,][]{sengupta10, marley11, dekok11, sanghavi18}. Photometric and spectroscopic observations of brown dwarfs and the directly imaged gas giants \citep[e.g.,][]{burgasser02, marois08, zhou16}  indicate the presence of clouds in their atmospheres. In the present investigation, therefore, we consider the cloudy atmospheres of the young gas giants in order to estimate the observable polarization.

\cite{sengupta09, sengupta10, marley11} presented theoretical models based on the spherical harmonic expansion (SHE) technique in order to estimate the detectable polarization of a fast-rotating substellar object with a certain oblateness. In order to estimate the disk integrated polarization, these authors followed the technique prescribed by \cite{simmons82}. They have presented polarization models for different properties of the substellar objects using self-consistent radiative-convective equilibrium models for the atmospheres. The model presented by \cite{jensen20} calculates polarization from the rotation-induced oblateness using the same technique but also includes evolution models to self-consistently calculate the moment of inertia, size, and surface gravity of a rotating substellar object at a given mass and age. \cite{dekok11} presented numerical models for the polarization observable from the directly imaged planets due to various sources of asymmetry such as oblateness, banded clouds, hot spots, etc. They followed a technique similar to that of \cite{sengupta09, sengupta10, marley11} to correlate the rotation-induced oblateness with the observable polarization. Again, \cite{sanghavi18, sanghavi19} included the factors such as gravitational darkening and flattening of the planetary disks in order to calculate the polarization from the fast-rotating brown dwarfs. On the other hand, \cite{stolker17} presented a Monte Carlo-based technique to calculate the polarization from a self-luminous exoplanet arising from its oblateness and due to the presence of banded and patchy clouds. However, none of these works consider the effect of the anisotropy in the atmosphere of the ellipsoidal planet across the disk. 
 
In this paper, we present a technique to calculate the polarization detectable from a substellar object with an oblate spheroid shape by incorporating the effects of both vertical (layerwise) inhomogeneity and horizontal (across the disk) anisotropy of the atmospheres of the planets. The atmosphere of an oblate planet is usually anisotropic inherently even if the gas molecules or the cloud particles across the disk are distributed homogeneously. This anisotropy in the atmosphere arises due to the fact that the depth of the atmospheric shell itself varies across the disk depending on its latitude and the inclination angle $i$ \citep{simmons82}. We calculate the disk-resolved and disk-integrated emergent radiation field by using the same numerical recipes described in \cite{chakrabarty21}. However, the atmospheric models and the radiative transfer equations that we solve for the present cases, are different since we focus on the thermal emission from the young wide-orbit giant exoplanets and not the reflected flux.  In order to calculate the layerwise and wavelength-dependent properties of the atmospheres of such Exoplanets, we use the state-of-the-art Sonora models \citep{marleyz18, marleyz21, marley21} which are available as grids of effective temperature ($T_{\rm eff}$), surface gravity ($g$), and metallicity. These calculations are based on the radiative-convective equilibrium and evolutionary models \citep{saumon08, marley11, jensen20} of the substellar objects. We have also incorporated a generalized Henyey-Greenstein-Rayleigh phase matrix for a better representation of the scattering due to the cloud particulates present in the atmospheres. This new technique will allow us to model the flux and polarization observable from the substellar objects with inhomogeneous or patchy atmospheres in our follow-up investigation.

\begin{figure}[!ht]
\centering
\includegraphics[scale=0.5,angle=0]{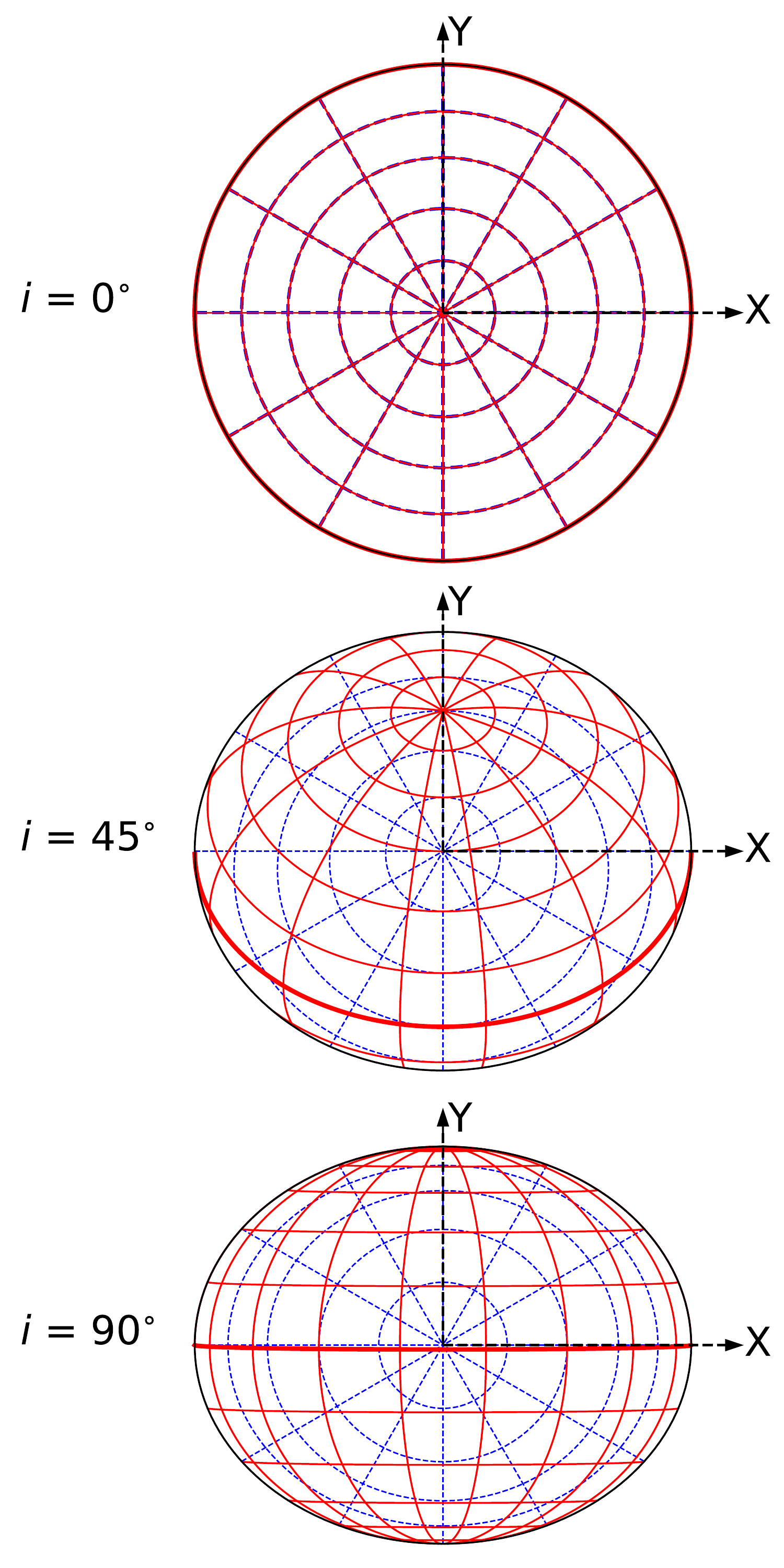}
\caption{The planetary disks are divided into grids, shown at three different inclination angles ($i$) of the rotation axis with respect to the observer. The grid of latitudes and longitudes are shown in red and the grid of disk-centered coordinates, i.e., $M$ and $\Phi$ are shown in blue. The red bold line denotes the equator of the planet. The Z-axis is assumed to be in the outward direction of the plane of paper which is also the direction towards the observer.
\label{fig:geom}}
\end{figure}

We have chosen the self-luminous directly imaged Exoplanets such as $\beta$ Pic b and ROXs 42B b whose projected rotation velocity $v_e\sin i$ are reported to be $25\pm3$ km/s \citep{snellen14} and $9.5_{-2.3}^{+2.1}$ km/s \citep{bryan18} respectively. Also, \cite{jensen20} reported the J-band polarimetric observations of these planets. They could not measure the polarization of these planets precisely but could only set upper limits on the observed polarization which can be attributed to these planetary sources. However, a number of future observations and dedicated missions are being planned for precise polarimetric characterization of the exoplanets. Hence, in the present study, we have developed models for the polarization detectable from these planets in four different wavelength bands of the SPHERE-IRDIS instrument to present the maximum amount of polarization that can be expected from these planets in those wavelength bands mainly due to their oblate structures. 

Section~\ref{sec:obl} shows the relation between the rotation rate and the induced oblateness of a fast-rotating planet as well as the effect of this oblateness on the shape of the atmosphere. Section~\ref{sec:calc}, which is further divided into four subsections, provides a detailed description of our overall approaches such as the atmospheric model adopted, the numerical technique used to calculate the local Stokes vectors, the detailed calculation of the Henyey-Greenstein-Rayleigh phase matrix in regard to the scattering by clouds, and the numerical recipe developed to calculate the disk-integrated flux and polarization from an oblate spheroid planet. We elaborate the specific band-averaged models that we have developed for the planets $\beta$ Pic b and ROSx 42 Bb for different values of $v_e\sin i$ and $i$ in Section~\ref{sec:betaroxobs}. The results are discussed in detail in Section~\ref{sec:rd} and the investigation is concluded in Section~\ref{sec:con}.

\section{Shape of a Fast-Rotating Planet} \label{sec:obl}

The oblateness of a fast-rotating young giant planet can be expressed as, $f = 1 - R_p/R_e$, where $R_p$ and $R_e$ are the polar and the equatorial radii of the planet respectively. The oblateness ($f$) depends on the rotation rate ($\Omega$), the surface gravity ($g$) and the mass of the planet (M$_{\rm P}$) as dictated by the Darwin-Radau relationship \citep{barnes03}, given by,

\begin{equation} \label{eq:dreq}
f = \frac{\Omega^2 R_e}{g}\left[\frac{5}{2}\left(1-\frac{3K}{2}\right)^2+\frac{2}{5}\right]^{-1},
\end{equation}

where $K$=$I$/(M$_{\rm P}R_e^2$), $I$ being the moment of inertia of the planet. As $I$ does not depend on the oblateness to the first order \citep{barnes03}, $I$ can be calculated by assuming the planet to be spherical. The moments of inertia of the self-luminous and cloud-free substellar objects can be obtained from the Sonora Bobcat tables \citep{marleyz21} available online\footnote{\url{https://doi.org/10.5281/zenodo.5063476}}. However, for most of our calculations for the cloudy self-luminous giant planets, we have assumed the interior of the planet to be a stable polytropic gas of index n=1 by following \cite{marley11} (henceforth, MS11)and hence set $K$=0.261 in Equation~\ref{eq:dreq} \citep{chandrasekhar33}. From the oblateness-rotation speed relationship (see Figure 15) of \cite{jensen20}, the value of $K$ for the planets $\beta$ Pic b and ROXs 42B b can be found out to be 0.276 and 0.297 respectively. Clearly, our assumption of $K$=0.261 provides a valid representation of the moments of inertia of the self-luminous giant planets. However, while calculating the models for those planets (see Section~\ref{sec:betaroxobs}) we have assumed the specific values of $K$ mentioned above.

The atmosphere of a fast-rotating planet can be considered to be an oblate spheroidal shell (see Figure 2b of \cite{simmons82}). The outermost radius of the planet at a colatitude $\Theta_{\rm col}$ across the disk can be expressed as $R(\Theta_{\rm col}) = r(\Theta_{\rm col})R_e$. The factor $r(\Theta_{\rm col})$ can be defined as \citep{sengupta09,simmons82},
\vspace{-1em}
\begin{equation} \label{eq:oblfac}
r(\Theta_{\rm col}) = \frac{1}{[1+(A^2-1)\cos^2\Theta_{\rm col}]^{1/2}},
\end{equation}

where $A=1/(1-f)=R_e/R_p$. Clearly, the shell has a varying thickness across the disk and the thickness at a colatitude $\Theta_{\rm col}$ is equal to $r(\Theta_{\rm col})$ times the thickness along the equatorial plane.

Observations \citep[e.g.,][]{snellen14} suggest that the young gas giants and the brown dwarfs do not have any systematic difference in their rotation rates. However, the relatively low surface gravity of the young gas giants allows them to attain higher rotation-induced oblateness, as high as $\sim0.2-0.3$, compared to the brown dwarfs. In all of our calculations, we consider the oblateness of the planets to be $\lesssim0.44$, as beyond this limit (this upper limit is for n=1; for higher values of n the upper limit is even lower), the atmospheres of the gas giant planets or the brown dwarfs are likely to become unstable  \citep[e.g.,][]{james64, marley11, sanghavi18}. 

\begin{figure}[!ht]
\centering
\includegraphics[scale=0.3,angle=0]{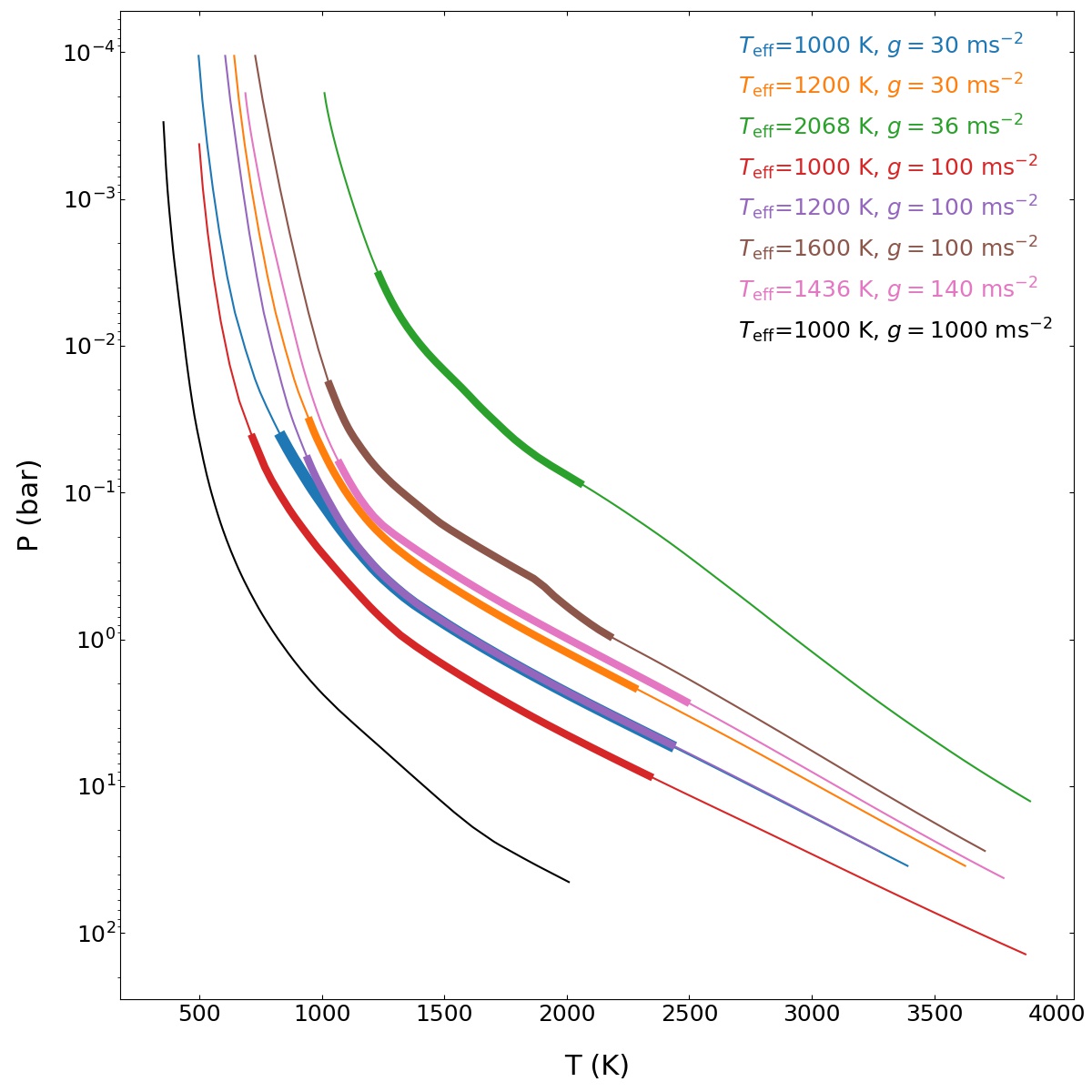}
\caption{The pressure-temperature profiles of the cloudy ($f_{\rm sed}=2$) directly imaged planets and the T-dwarf described in Section~\ref{sec:calc/atm}. The thick regions denote the cloudy layers between the cloud base and cloud deck of the planets.
\label{fig:pt}}
\end{figure}

\section{Calculation of Flux and Polarization} \label{sec:calc}

\subsection{Model atmosphere of self-luminous Exoplanets} \label{sec:calc/atm}

In the present work, we use the Sonora grid of pressure-temperature (P-T) profiles \citep[e.g.,][]{marley21, jensen20, marley11, sengupta10, sengupta09, saumon08, marley02, ackerman01} calculated for the atmospheres of the substellar mass objects by using radiative-convective equilibrium, chemical equilibrium, and evolutionary models self-consistently. We also use the pre-calculated Sonora grids for the layerwise properties of the atmospheres such as mass extinction coefficients, mass absorption coefficients, and the scattering asymmetry parameters. These models are based on the opacity calculations of \cite{freedman08, freedman14} and the molecular abundance calculations of \cite{lodders20, lodders10}. In all of our derivations, we assume solar metallicity and solar-system value for the Carbon-to-Oxygen (C/O) ratio. Also, we assume the pressure and temperature to be uniform along each of the stratified oblate spheroid layers, calculated over a range of effective temperature ($T_{\rm eff}$) and surface gravity ($g$) that describe the atmospheres of the young gas giants. Figure~\ref{fig:pt} shows the P-T profiles of the young gas giants for different values of $T_{\rm eff}$ and $g$.

To start with, we consider the cloud-free atmosphere of a Jupiter-sized T brown dwarf or methane dwarf with $T_{\rm eff}=1000$ K and $g=1000$ ms$^{-2}$ to benchmark our calculations based on Rayleigh scattering. We use the corresponding P-T profile (see Figure~\ref{fig:pt}), mass extinction coefficients, and mass absorption coefficients from the Sonora Bobcat model for cloud-free atmospheres \citep{marley21, marleyz21, marleyz18, sengupta09}. We consider an equatorial speed ($v_e$) of 90 kms$^{-1}$ that corresponds to a rotation period ($P_{\rm rot}$) of 1.386 hour and $i=90^\circ$. We then compare our results with the results presented by \cite{sengupta09}.

Next, we use the cloudy Sonora models to calculate the polarization detectable from the young self-luminous giant planets \citep{jensen20, marley11}. We consider young planets with $T_{\rm eff}$ ranging between 1000 K and 2068 K and $g$ between 10 ms$^{-2}$ and 140 ms$^{-2}$. For the cloud models, we assume a sedimentation efficiency, $f_{\rm sed}=2$ \citep{ackerman01}. In order to solve the vector radiative transfer equations, we apply $\delta$-Eddington approximation \citep{joseph76, batalha19} on the optical depth, single-scattering albedo, and asymmetry parameter before calculating the Henyey-Greenstein-Rayleigh phase matrix for the cloud particles.

\subsection{Calculations of the Stokes vectors at local points of the atmosphere} \label{sec:calc/intvec}

Similar to the case of brown dwarfs \citep{sengupta09, sengupta10}, MS11 use the SHE technique to calculate the local intensity vectors (or, Stokes vectors) across the disk of the self-luminous gas giant exoplanets. In order to account for the effect of the oblateness on the polarization of these planets, they follow the formalism presented by \cite{simmons82} which has also been explained in detail by \cite{sengupta09}. While the formalism prescribed by \cite{simmons82} is a first-order approximation and uses less computational effort,  a better and more accurate approach is needed in order to calculate the disk-integrated polarization arising purely due to the oblateness of the planets. We developed a new technique to incorporate the effect of the anisotropy of the medium over the disk by using a three-dimensional approach. 

As explained in \cite{chakrabarty21}, we divide the observable planetary disk into a grid of $M$ and $\Phi$, where $M$ denotes the cosine of the polar angular position and $\Phi$ denotes the azimuthal angular position on the disk defined with respect to the disk center $M=1$. We consider $\Phi=0$ along the Y-axis. The rotation axis of the planet is assumed to lie along the YZ-plane (see Figure~\ref{fig:geom}) depending on the inclination angle of the rotation axis, $i$ , with respect to the observer. When $i=90^\circ$, the poles lie on the Y-axis, and for $i=0^\circ$, the poles lie on the Z-axis i.e., on the line of sight (LOS) of the observer. Clearly, when $i=0^\circ$ (polar view), the disk exhibits circular symmetry (uniform along $\Phi$), and hence, the net disk-integrated polarization becomes zero in this case. On the other hand, the effect of the oblateness is maximum for $i=90^\circ$, i.e. for equatorial view. We transform the disk-centric coordinates ($M,\Phi)$ to the corresponding colatitudes ($\Theta_{\rm col}$) of the planet to calculate the effect of the oblateness of the planet as shown in Equation~\ref{eq:dreq}-\ref{eq:oblfac}. For an inclination angle, $i$, we calculate the colatitudes by,

\begin{equation} \label{eq:mphi2lat}
\cos\Theta_{\rm col} = M\cos i + \sqrt{1-M^2}\sin i \cos\Phi.
\end{equation}

Unlike the case presented by \cite{chakrabarty21}, here we assume the planetary atmospheric shell to be stratified and homogeneous along the concentric oblate spheroid layers. We divide the direction of propagation of light at each location on the disk into a grid of $\mu$ and $\phi$, where $\mu$ denotes the cosine of the polar angular direction and $\phi$ denotes the azimuthal angular direction defined with respect to the local meridian. While solving the radiative transfer equations for the case of thermal emission, we consider the radiation field to be circularly symmetric about the local normal at each location on the disk. However, in the present study, we assume circular symmetry about the radial line joining the center of the planet and the point on the disk in order to avoid further complications, leveraging the fact that $\omega<0.44$. As a result, we can assume the incoming and outgoing radiation to be uniform along the $\phi$-direction. Hence, we adopt the axisymmetric form of the vector radiative transfer equation including multiple scattering of the internal radiations at each location ($M, \Phi$) and at each layer with an optical depth, $\tau(M,\Phi)$ defined from the top of atmosphere (TOA). This is given by,

\begin{equation} \label{eq:vrt}
\mu\frac{d\vec{I_{\rm mer}}(\tau,M,\Phi,\mu)}{d\tau} = \vec{I_{\rm mer}}(\tau,M,\Phi,\mu)
-\frac{\omega(\tau)}{2}\int_{-1}^1 \vec{P_{\rm mer}}(\mu;\mu')\vec{I_{\rm mer}}(\tau,M,\Phi,\mu') ~d\mu' - (1-\omega)\vec{B}(\tau)
\end{equation}

The wavelength-dependent intensity vector (or, Stokes vector) $\vec{I_{mer}}=[I_{\rm mer}, Q_{\rm mer}, U_{\rm mer}, V_{\rm mer}]$ (see, e.g., \cite{chakrabarty21, chandrasekhar60}) is a function of the optical depth ($\tau$), disk location ($M,\Phi$) and direction ($\mu$). The subscript `mer' signifies that this vector is defined with respect to the local meridian plane at ($M,\Phi$). \mvec{B}($\tau$) denotes the unpolarized internal radiation source function from a layer with optical depth, $\tau$ from TOA. Therefore, $\vec{B}(\tau)=[B(\tau), 0, 0, 0]$. $\omega$ denotes the single-scattering albedo of the atmosphere at $\tau$. 

For linear polarization due to scattering, $V_{\rm mer}=0$.  $P_{\rm mer}(\mu;\mu')$ denotes the $\phi$-averaged phase matrix from the direction $\mu'$ to the direction $\mu$. We consider the Rayleigh phase matrix (see Equation~5 of \cite{chakrabarty21}) for the cloud-free atmosphere and for a cloudy atmosphere, we add the Henyey-Greenstein-Rayleigh (HGR) phase matrix \citep{liu06} to the Rayleigh phase matrix in the ratio of their single-scattering albedos, $\omega_{cld}$ and $\omega_{mol}$ respectively \citep[MS11,][]{chakrabarty21}. In our axisymmetric calculations, both the Rayleigh and HGR phase matrices do not contribute to the $U$-component and hence, we can set $U_{\rm mer}=0$. Further detail on the calculations of HGR phase matrix is described in Section~\ref{sec:calc/hgr}.

We solve Equation~\ref{eq:vrt} using Discrete Space Theory \citep[e.g.,][]{chakrabarty21, chakrabarty20, sengupta20, marley11, sengupta10, sengupta09} and finally, calculate the intensity vector at the TOA towards the observer by,

\begin{equation} \label{eq:imerobs}
\vec{I_{\rm mer, obs}(M,\Phi)} = \vec{I_{\rm mer}}(\tau=0,M,\Phi,\mu=M)
\end{equation}

For validity check,  we first adopt the SHE formalism used by \cite{sengupta09} and compare the results with that calculated by using the new technique developed and described in this paper.  For this purpose, we considered the case for a  cloud-free atmosphere of a T-dwarf with $T_{\rm eff}=1000$ K and $g=1000$ ms$^{-2}$  and the case for a cloudy atmosphere of a young giant planet with $T_{\rm eff}=1000$ K and $g=30$ ms$^{-2}$. We first consider the object to be perfectly spherical while calculating the intensity vectors ($\vec{I_{\rm mer, obs}(M,\Phi)}$). The flux and the polarization calculated at all the local points of the planetary disk are then integrated over an oblate spheroid by using the SHE formalism. This is done by including the factor $r(\Theta_{\rm col})$ as explained in \cite{sengupta09}.

Next, in order to account for the effect of the anisotropy of the atmosphere due to the rotation-induced non-sphericity of the object,  we introduce the optical depth ($d\tau$) at each stratified layer as a function of the colatitude ($\Theta_{\rm col}$) and hence, a function of $M$ and $\Phi$ (see Equation~\ref{eq:mphi2lat}), which is given by,

\begin{equation} \label{eq:ellipdepth}
d\tau(\Theta_{\rm col}(M,\Phi)) = r(\Theta_{\rm col}(M,\Phi))~d\tau_e,
\end{equation}
where $d\tau_e$ = $d\tau(\Theta_{\rm col}=0)$, which is the optical depth along the equatorial plane. By substituting Equation~\ref{eq:ellipdepth} into Equation~\ref{eq:vrt} we calculate $\vec{I_{\rm mer, obs}(M,\Phi)}$ at each location on the disk. Thus in our present formalism, the optical depth is considered to be anisotropic. 

Since $\vec{I_{\rm mer, obs}(M,\Phi)}$ is defined with respect to the local meridian at each location on the disk, we need to transform $\vec{I_{\rm mer, obs}(M,\Phi)}$ to a common plane of reference all over the disk before integrating them.
We choose the YZ-plane as this common plane of reference. Accrodingly, we apply the rotation matrix, $\vec{L}(\Phi)$ (See Equation~2 of \cite{chakrabarty21}) to calculate the intenisty vector $\vec{I_{\rm obs}(M,\Phi)}$ defined with respect to the YZ-plane as,

\begin{equation} \label{eq:iobs}
\vec{I_{\rm obs}(M,\Phi)} = \vec{L}(\Phi)~\vec{I_{\rm mer, obs}(M,\Phi)}
\end{equation}

From Equation~\ref{eq:iobs}, we calculate the components of $\vec{I_{\rm obs}(M,\Phi)}$ as,

\begin{equation} \label{eq:iobs1}
\begin{aligned}
I_{\rm obs}(M,\Phi) & = I_{\rm mer, obs}(M,\Phi)\\
Q_{\rm obs}(M,\Phi) & = Q_{\rm mer, obs}(M,\Phi) \cos 2\Phi\\
U_{\rm obs}(M,\Phi) & = -~\!Q_{\rm mer, obs}(M,\Phi) \sin 2\Phi
\end{aligned}
\end{equation}

\begin{figure}[!ht]
\centering
\includegraphics[scale=0.3,angle=0]{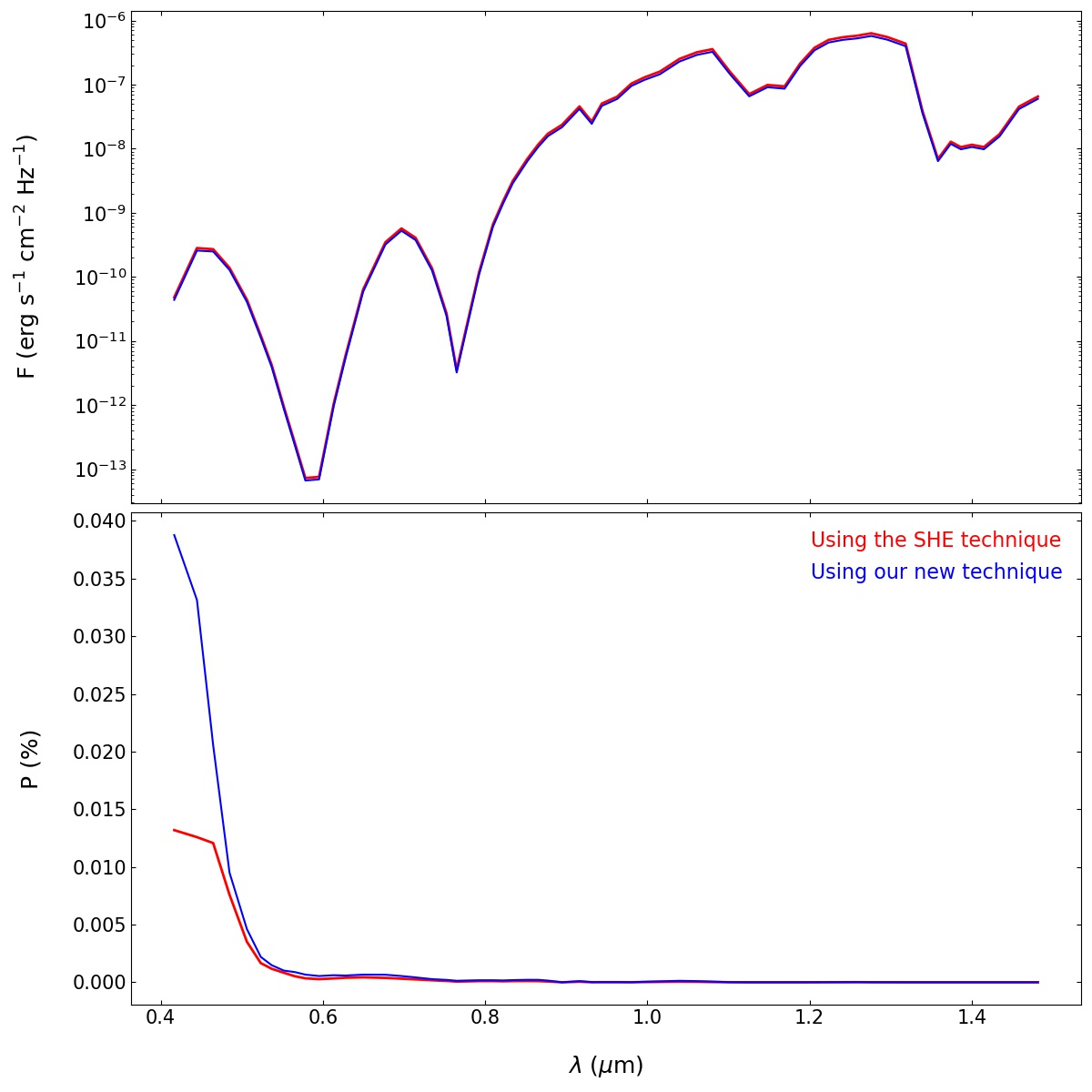}
\caption{The flux and polarization spectra of a cloud-free T-dwarf with $T_{\rm eff}=1000$ K and $g=1000$ ms$^{-2}$ rotating at a period of 5 h, calculated by following the spherical harmonic expansion (SHE) method of \cite{sengupta09} and by using the newly developed technique presented in Section~\ref{sec:calc}.
\label{fig:tdwarf}}
\end{figure}

\begin{figure}[!ht]
\centering
\includegraphics[scale=0.4,angle=0]{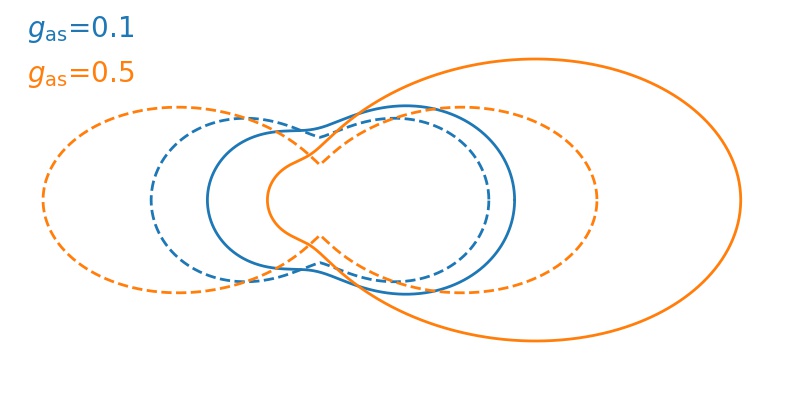}
\caption{The axisymmetric ($\phi$-averaged) HGR phase functions calculated following MS11 (dashed) and applying the modification explained in Section~\ref{sec:calc/hgr} (solid) which present the significance of the difference between the two approaches at higher values of $g_{\rm as}$.
\label{fig:phasefunc}}
\end{figure}

\begin{figure}[!ht]
\centering
\includegraphics[scale=0.35,angle=0]{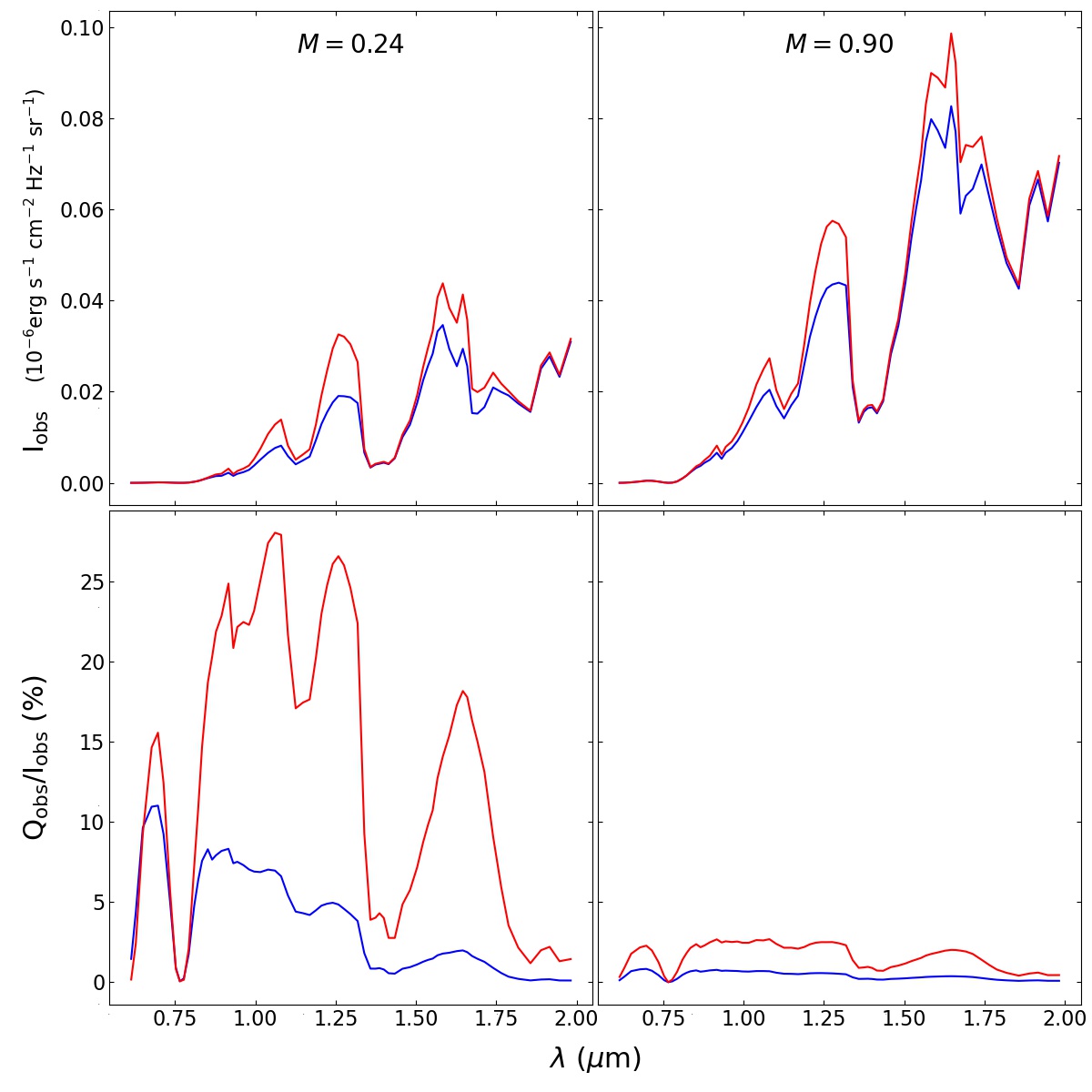}
\caption{The effect of the scattering phase matrix of the cloud particles on the total local intensity ($I_{\rm obs}$) and the local degree of polarization ($Q_{\rm obs}/I_{\rm obs}$) at two different polar angular positions on the disk of a self-luminous giant planet with $T_{\rm eff}=1000$ K and $g=30$ ms$^{-2}$. The calculations are performed by adopting the HGR phase function (red) of MS11 and the generalized HGR phase function (blue) as explained in Section~\ref{sec:calc/hgr}, ignoring the effect of oblateness. 
\label{fig:pfonintvec}}
\end{figure}

\begin{figure}[!ht]
\centering
\includegraphics[scale=0.35,angle=0]{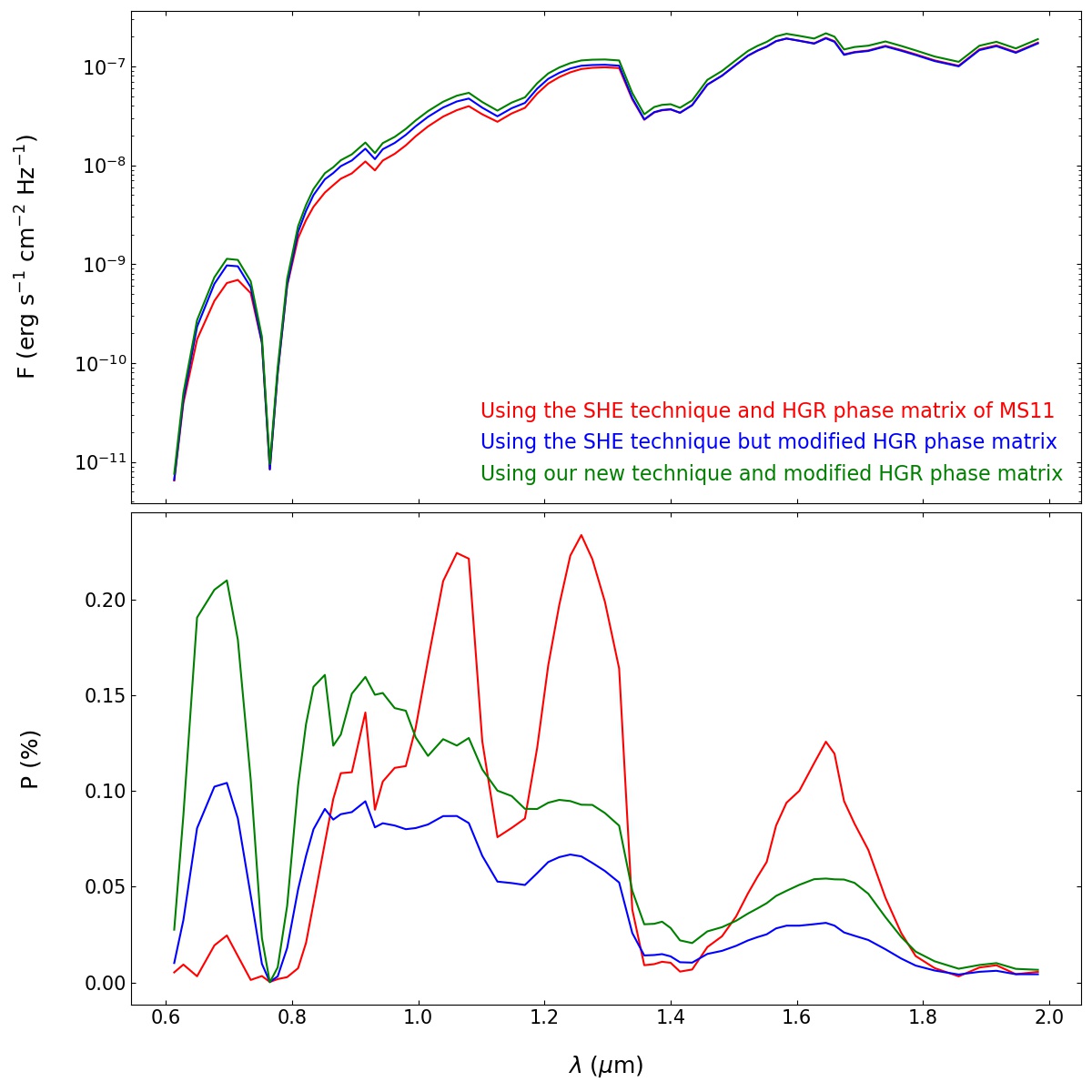}
\caption{The flux and polarization spectra of a cloudy young oblate self-luminous planet with $T_{\rm eff}=1000$ K and $g=30$ ms$^{-2}$ rotating at a period of 5 h, calculated by using the spherical harmonic expansion (SHE) method of \citep{sengupta09} and by using the newly developed technique presented in Section~\ref{sec:calc}. The flux and polarization calculated with and without the generalized HGR phase matrix (see MS11 and Section~\ref{sec:calc/hgr}) are also presented.
\label{fig:dip}}
\end{figure}

\begin{figure}[!ht]
\centering
\includegraphics[scale=0.5,angle=0]{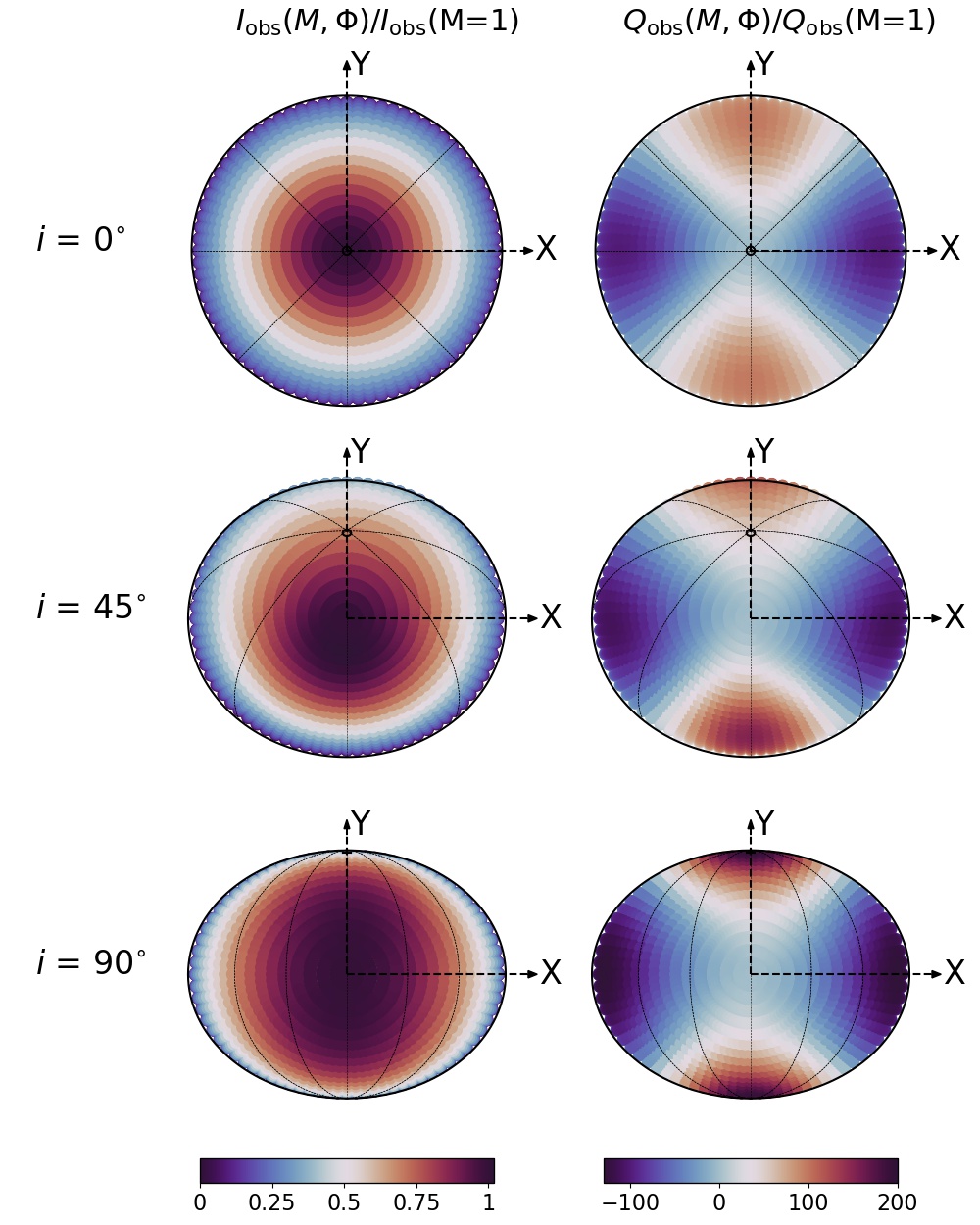}
\caption{The color maps of total intensity ($I_{\rm obs}$) and the polarized intensity ($Q_{\rm obs}$) over the disk in the direction of the observer normalized with respect to the values at the disk center of a cloudy young oblate self-luminous planet with $T_{\rm eff}=1000$ K, $g=30$ ms$^{-2}$, and $P_{\rm rot}=5$ h. $I_{\rm obs}$ exhibits limb darkening whereas $Q_{\rm obs}$ exhibits limb brightening. The limb polarization dominates over the polarization from the disk center. At $i=90^\circ$, the disk shows maximum asymmetry of $Q_{\rm obs}$ over $\Phi$ causing maximum disk-integrated polarization at $i=90^\circ$ for a given rotation period.
\label{fig:contour5}}
\end{figure}

\begin{figure}[!ht]
\centering
\includegraphics[scale=0.5,angle=0]{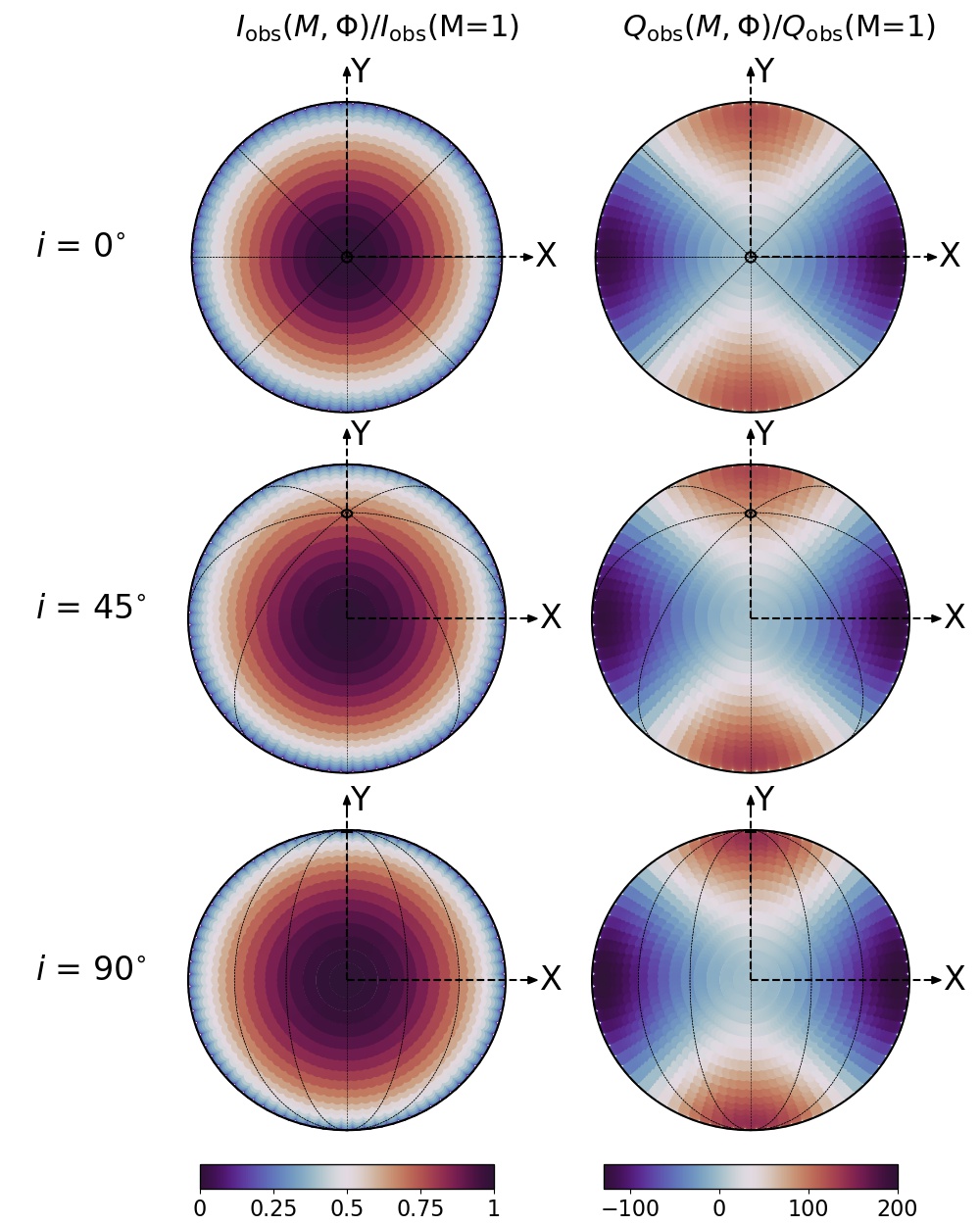}
\caption{Same as Figure~\ref{fig:contour5} but for $P_{\rm rot}=10$ h.
\label{fig:contour10}}
\end{figure}

\subsection{The Generalized Henyey-Greenstein-Rayleigh phase matrix} \label{sec:calc/hgr}

The phase matrix $P_{\rm mer}$ that determines the angular distribution of photons before and after scattering, strongly dictates the amount of polarization. In the case of cloudy self-luminous Exoplanets, MS11 (also see \cite{sengupta10,jensen20}) have used Henyey-Greenstein-Rayleigh (HGR) phase matrix \citep{liu06} where they have assumed a Rayleigh-dominated symmetric form for the phase function (See Figure~\ref{fig:phasefunc}). This is valid for the axisymmetric radiation field. However, this approach tends to overestimate the polarization for higher values of the asymmetry parameter $g_{\rm as}$ because the amount of light scattered in the forward and backward directions differs greatly at higher values of $g_{\rm as}$ as shown in Figure~\ref{fig:phasefunc}. For this reason, we have used the azimuth-dependent expression of the HGR phase matrix as given in Equation 9 of \cite{liu06}. Therefore, in order to derive the azimuth-dependent phase matrix $\vec{P_{\rm HGR}}(\mu,\phi;\mu',\phi'; g_{\rm as})$, we replace the term $\cos \Theta$ by 

\begin{equation} \label{eq:scatangle2d}
\cos\Theta = \mu\mu' + \sqrt{(1-\mu^2)(1-\mu'^2)}\cos~(\phi'-\phi),
\end{equation}
where $\phi'$ and $\phi$ are the azimuthal angles that determines the directions of the incoming (before scattering) and the outgoing (after scattering) photon respectively. However, we have adjusted for the change in the reference planes of the intensity vectors during a scattering process from the direction ($\mu',\phi'$) to ($\mu,\phi$), following \cite{chakrabarty21} 

\begin{equation}
\vec{P_{\rm mer,HGR}}(\mu,\phi;\mu',\phi'; g_{\rm as}) = \vec{L}(-(\pi-i_2)) \vec{P_{\rm HGR}}(\mu,\phi;\mu',\phi'; g_{\rm as}) \vec{L}(i_1),
\end{equation}
where $i_1$ denotes the angle between the local meridian plane along \muphip ~and the plane of scattering and $i_2$ denotes the angle between the plane of scattering and the local meridian plane along \muphi. $\vec{P_{\rm mer,HGR}}$ denotes the HGR phase matrix defined with respect to the local meridian at \Mphi. The axisymmetric form of the phase matrix, i.e. $\vec{P_{\rm mer,HGR}}(\mu;\mu';g_{\rm as})$ is calculated by averaging $\vec{P_{\rm mer,HGR}}(\mu,\phi;\mu',\phi'; g_{\rm as})$ over $\phi'$ numerically.

Figure~\ref{fig:pfonintvec} shows the spectra, at a local point, of the intensity $I_{\rm obs}$ and the degree of polarization $Q_{\rm obs}/I_{\rm obs}$ in the direction towards the observer which are calculated by assuming the planet as a perfectly spherical body. It demonstrates the sole effect of the modified HGR phase matrix on the local intensity vector. The intensity and polarization are shown at two different polar angular points, viz. $M=0.24$ and $M=0.9$ by adopting the HGR phase matrix of MS11 and the generalized phase matrix.

\begin{figure}[!ht]
\centering
\includegraphics[scale=0.4,angle=0]{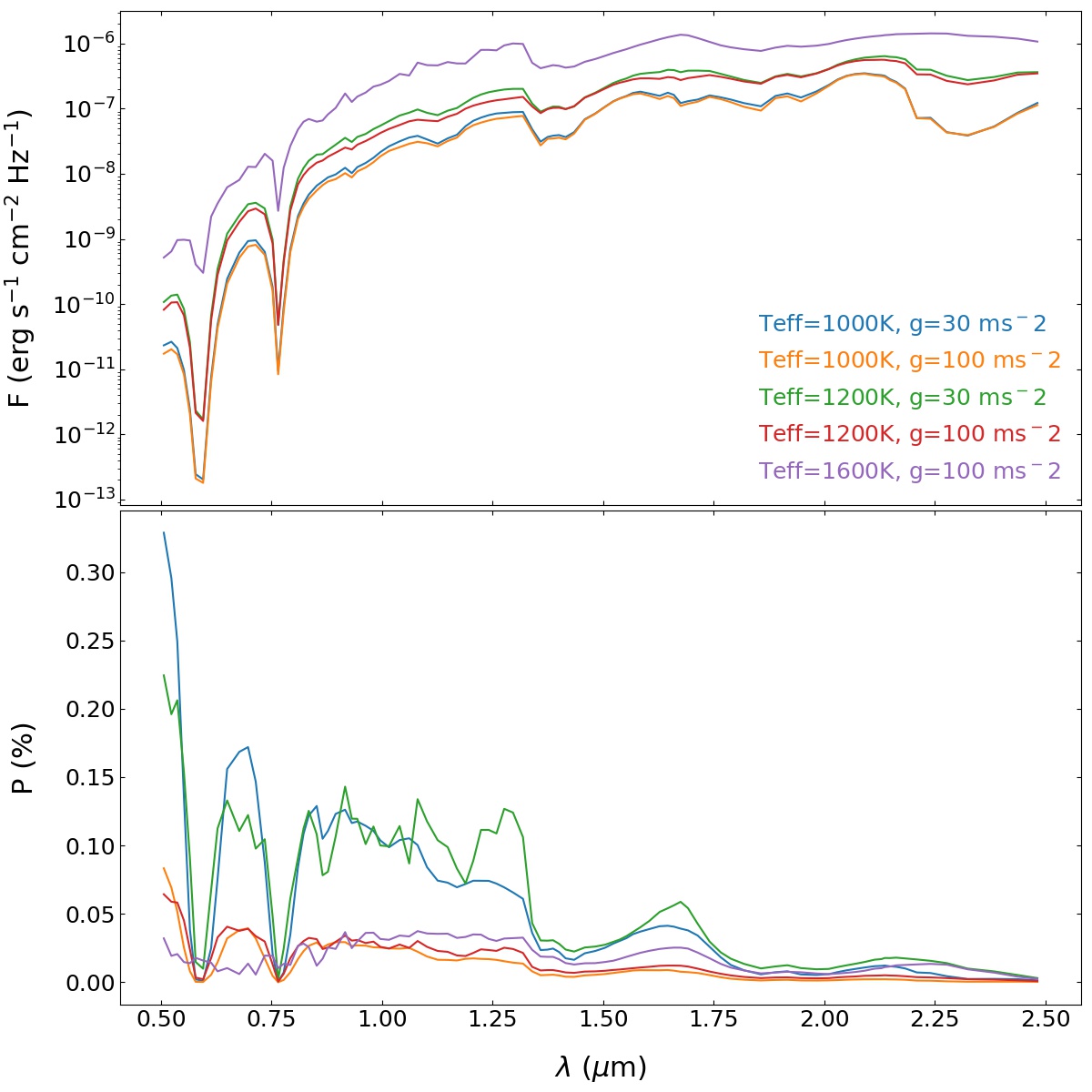}
\caption{The disk-averaged flux and polarization of an oblate and cloudy self-luminous Jupiter-sized planet with different values of $T_{\rm eff}$ and $g$.
 \label{fig:dip-teff-g}}
\end{figure}

\begin{figure}[!ht]
\centering
\includegraphics[scale=0.4,angle=0]{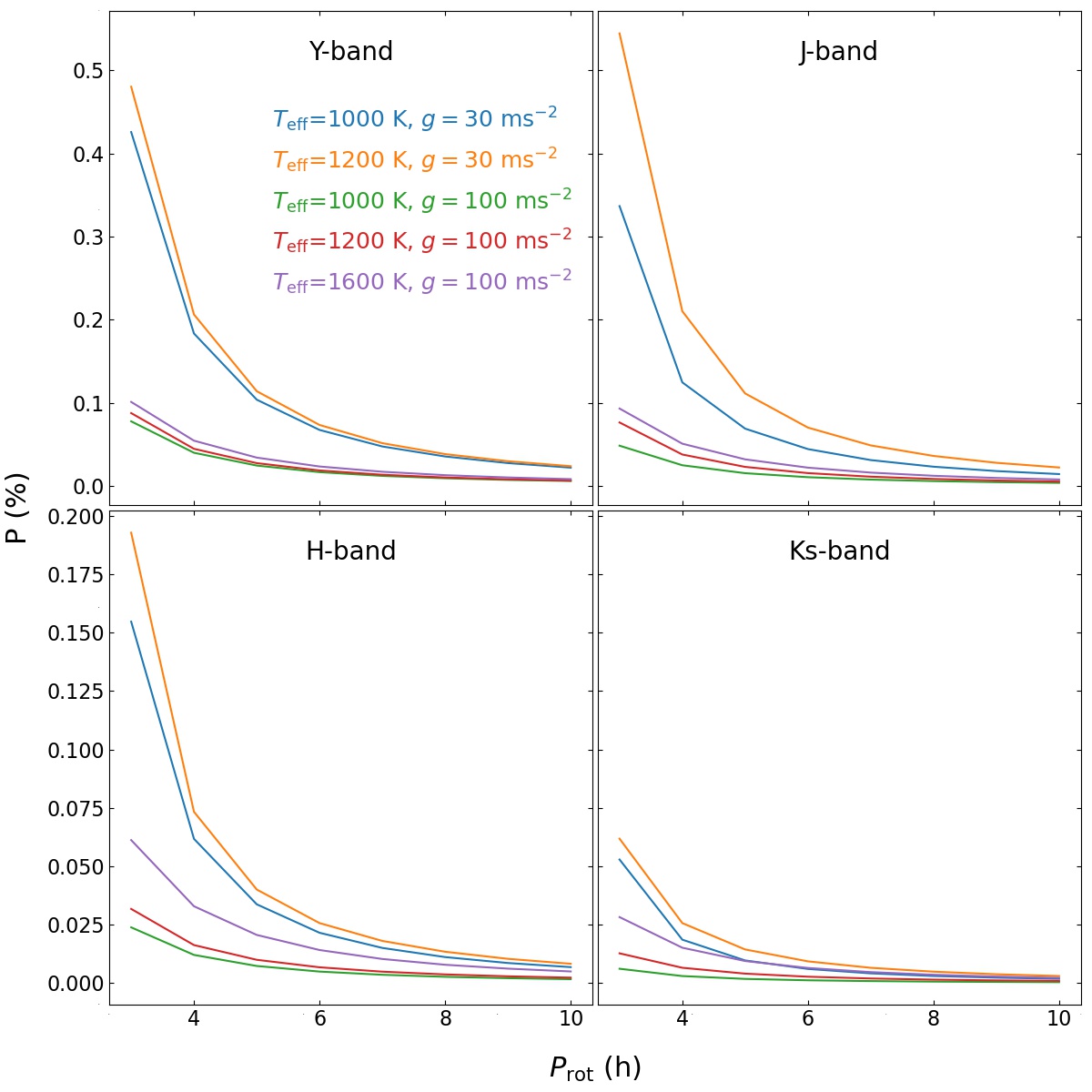}
\caption{The disk-averaged polarization as a function of rotation period $P_{\rm rot}$ at different wavelength bands of the SPHERE-IRDIS instrument for an oblate and self-luminous Jupiter-sized planet with different values $T_{\rm eff}$ and $g$. The inclination angle of the spin rotation axis is fixed at $i=90^\circ$. 
\label{fig:photpol-teffg}}
\end{figure}

\begin{figure}[!ht]
\centering
\includegraphics[scale=0.4,angle=0]{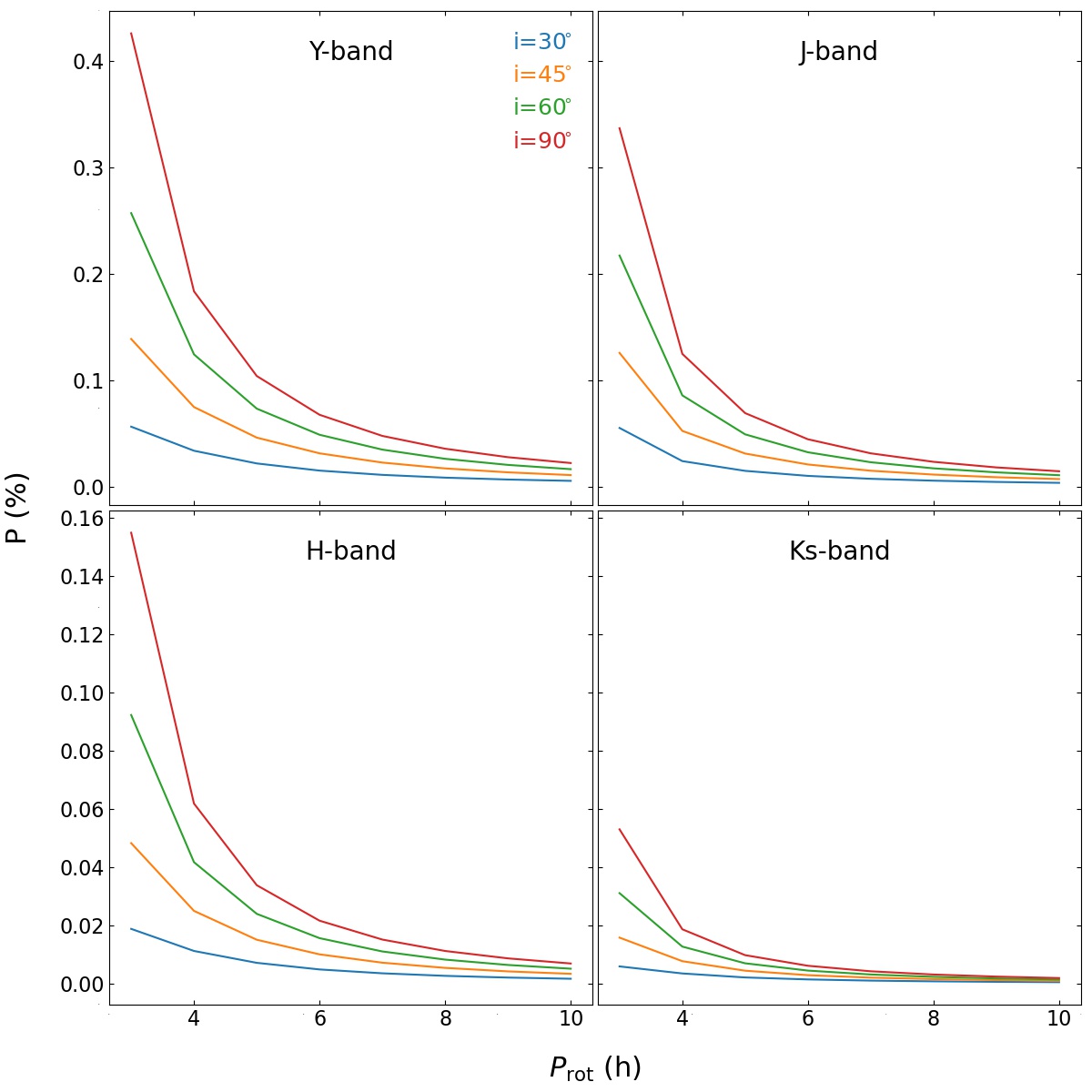}
\caption{Same as Figure~\ref{fig:photpol-teffg} but for different values of $i$.  The values of $T_{\rm eff}$ and $g$ are fixed at 1000 K and 30 ms$^{-2}$ respectively. 
\label{fig:photpol-inc}}
\end{figure}

\subsection{Integration of the local Stokes vectors over the oblate planetary disk} \label{sec:calc/diskint}

By integrating the intensity vector $\vec{I_{\rm obs}}$ over the solid angles of the disk projected to the observer, we get the disk-integrated flux vector $\vec{F} = [F, F_{\rm Q}, F_{\rm U}, F_{\rm V}]$, defined with respect to the YZ-plane. Here, $F$ denotes the total disk-integrated flux while $F_{\rm Q}$ and $F_{\rm U}$ denote the polarized disk-integrated flux ($F_{\rm V}]=0$ for linear polarization). $F_{\rm U}$ is also zero because of axisymmetry (see Equation~\ref{eq:iobs1}) with respect to the rotation axis. The disk-integrated degree of polarization is then given by,

\begin{equation} \label{eq:pdisk}
P = \frac{F_{\rm Q}}{F}
\end{equation}'

When we use the SHE method, we employ Equations 46 and Equation 47 of \cite{sengupta09} to calculate $F$ and $F_{\rm Q}$. On the other hand, when we apply the present technique, we calculate $F$ and $F_{\rm Q}$ by using the following expression: 

\begin{equation} \label{eq:fluxellipdepth}
\vec{F} = \frac{R_e^2}{D^2}\int_0^{2\pi}\int_{-1}^1 \vec{I_{\rm obs}}(M,\Phi)~\!M dM d\Phi,
\end{equation}
where D is the distance between the planet and the observer. We consider $\frac{R_e^2}{D^2}=1$ in Equation~\ref{eq:fluxellipdepth} so that the emergent flux and its state of polarization are estimated at the TOA of the planet. 

Figure~\ref{fig:tdwarf} shows the disk-integrated flux $F$ and polarization $P$ estimated by using both the SHE method \citep{sengupta09} and the technique described here (i.e., Equation~\ref{eq:fluxellipdepth}). On the other hand, Figure~\ref{fig:dip} shows a comparison of $F$ and $P$ derived by using the formalism used by MS11 and by using the present technique with and without the modified HGR phase matrix (see Section~\ref{sec:calc/hgr}).

 All further derivations are done by solving
 Equation~\ref{eq:fluxellipdepth} and employing the generalized HGR phase matrix. The corresponding color maps of the intensity and the polarization [$I_{\rm obs}(M, \Phi)$ and $Q_{\rm obs}(M, \Phi)$] at all points accorss the disk  of a self-luminous giant planet with $T_{\rm eff}=1000$ K and $g=30$ ms$^{-2}$ for two different rotation period $P_{\rm rot}=5$ h and $P_{\rm rot}=10$ h are shown in Figures~\ref{fig:contour5} and \ref{fig:contour10} respectively. These figures help us to visualize the asymmetry in the values of $Q_{\rm obs}(M, \Phi)$ over the $\Phi$-direction that gives rise to the net non-zero disk-integrated polarization the amount of which depends on $i$ and on the oblateness and hence on the rotation rate.  Similarly, Figure~\ref{fig:dip-teff-g} presents the detectable $F$ and $P$ of a self-luminous Jupiter-sized gas giant for $i=90^\circ$ and for different values of $T_{eff}$ and $g$. The corresponding atmospheric models are explained in Section~\ref{sec:calc/atm}. Figure~\ref{fig:photpol-teffg}-\ref{fig:photpol-inc} show the $P$ at different wavelength bands as a function of $P_{\rm rot}$ for planets with different values of $T_{\rm eff}$ and $g$ and different values of $i$ respectively. In order to calculate the flux and polarization at different wavelength bands, we have chosen the response functions of the broadband filters Y, J, H and K of
 SPHERE-IRDIS instrument placed at the backend of the Very Large Telescope (VLT).

\begin{figure}[!ht]
\centering
\includegraphics[scale=0.35,angle=0]{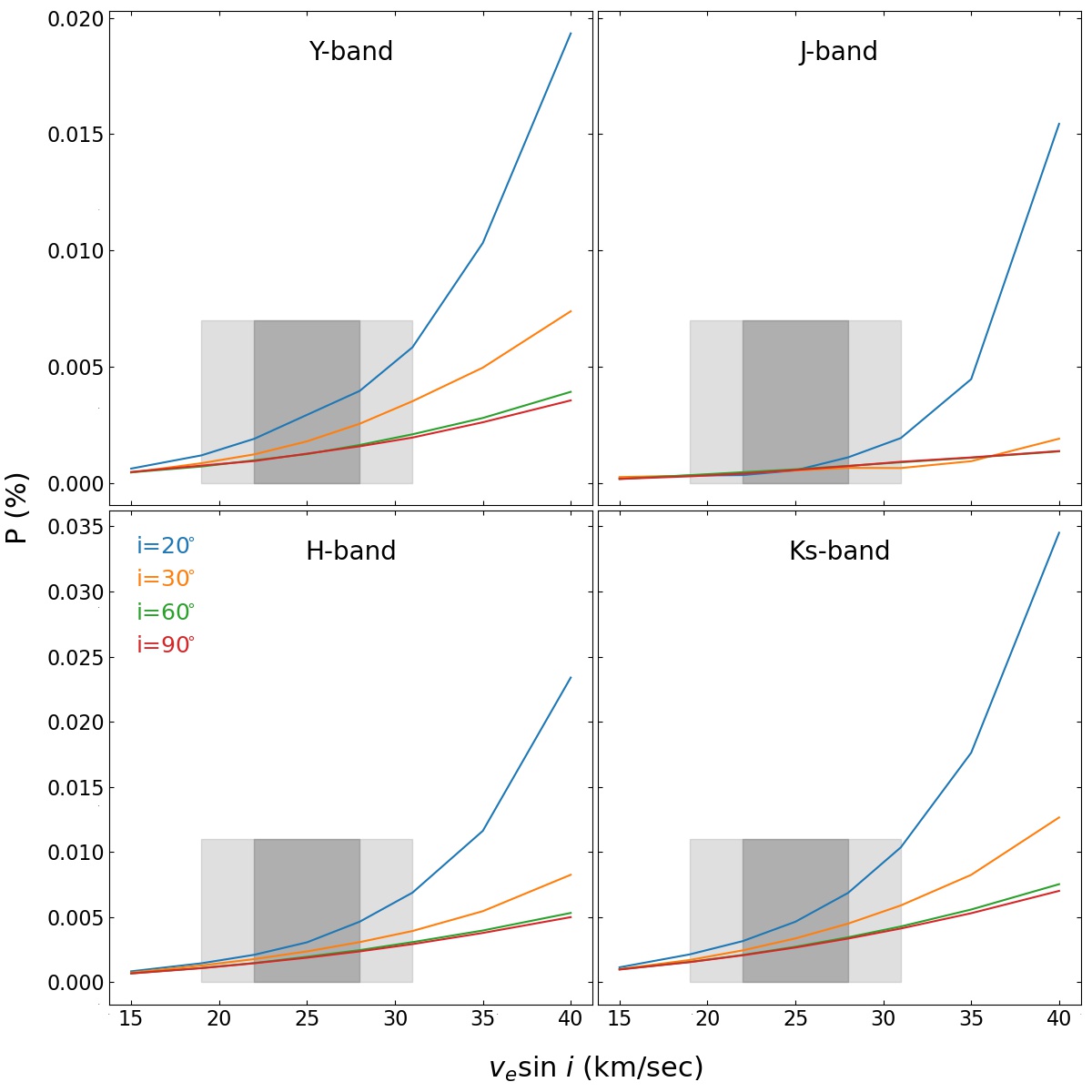}
\caption{The polarimetric models for the planet $\beta$ Pic b owing to its oblateness as functions of $v_e\sin i$ for different values of $i$ at different wavelength bands. The minimum value of $i$ is set from the given value of $v_e\sin i$ reported by \cite{snellen14} such that the rotation speed  $v_e$ does not exceed the stability limit of the planet. The dark and light shaded regions denote the 1-$\sigma$ and 2-$\sigma$ uncertainty domains of the estimated values of $v_e\sin i$ \citep{snellen14} respectively.}
\label{fig:betapic}
\end{figure}

\begin{figure}[!ht]
\centering
\includegraphics[scale=0.35,angle=0]{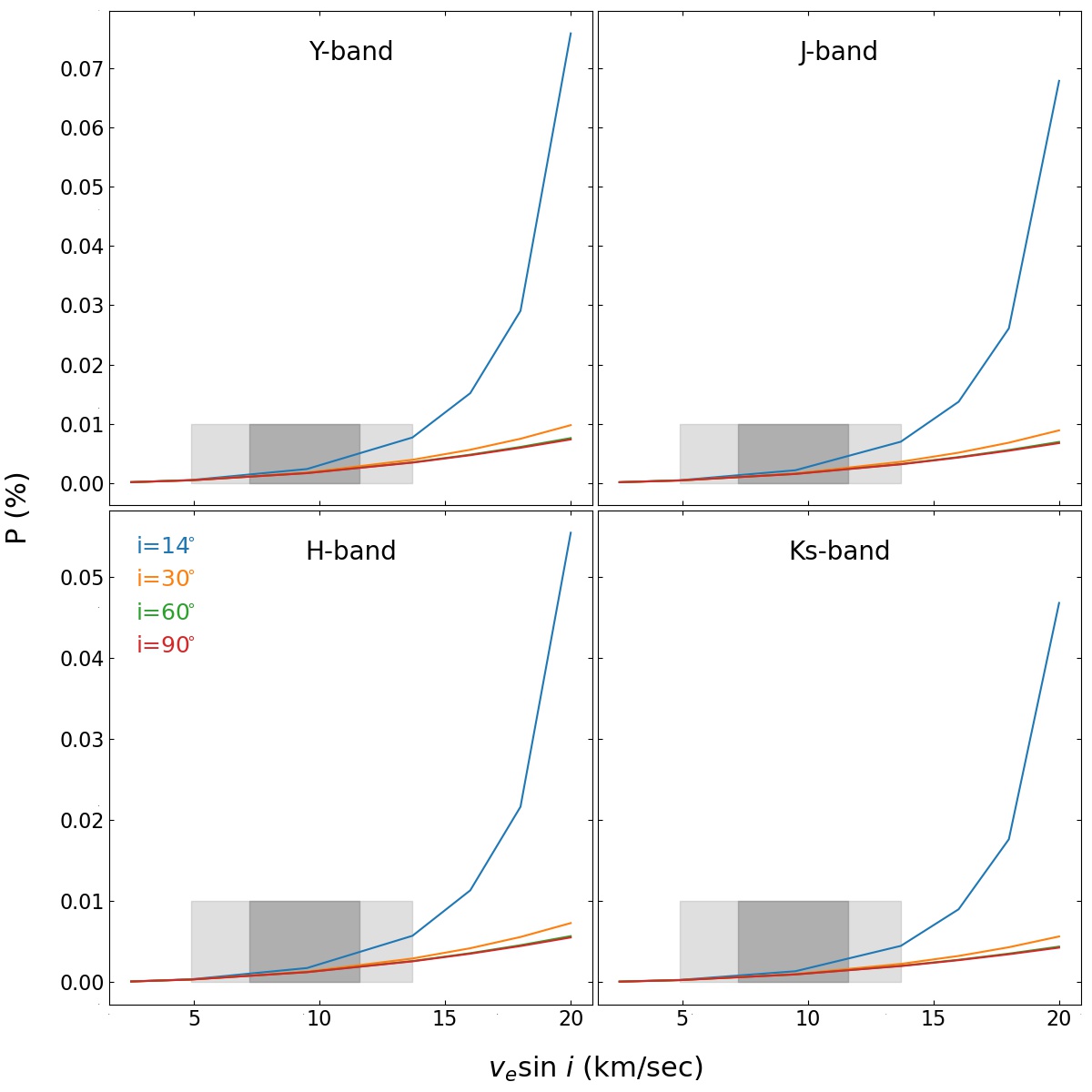}
\caption{Same as Figure \ref{fig:betapic} but for the planet ROXs 42B b with the shaded regions showing the uncertainty range of $v_e\sin i$ estimated by \cite{bryan18}.
\label{fig:roxs}}
\end{figure}

\section{Polarization of $\beta$ Pic \lowercase{b} and ROX\lowercase{s} 42B \lowercase{b}} \label{sec:betaroxobs}

High-resolution spectroscopic observations have been used to estimate the LOS components of the equatorial rotation velocities ($v_e\sin i$) of the young directly imaged planets such as $\beta$ Pic b \citep{snellen14}, ROXs 42B b \cite{bryan18}, among others. However, these studies cannot provide us with any information about the inclination angle of the rotation axis with respect to the observer. \cite{jensen20} have reported the polarimetric observation of these two planets in the J-band along with that of five brown dwarfs. Polarimetric observations of $\beta$ Pic b and ROXs 42B b have only been able to set upper limits on the degree of linear polarization.

 In Figures~\ref{fig:betapic} and \ref{fig:roxs} we present updated models for the polarization of the planets $\beta$ Pic b and ROXs 42B b respectively as functions of $v_e\sin i$ for different values of $i$. The oblatenesses for different equatorial rotation speeds have been calculated following \cite{jensen20} (see Figure 15). The amounts of polarization have been constrained by their observed values of $v_e\sin i$. In order to estimate the degree of polarization at different wavelength bands, we consider the response functions of the broadband filters corresponding to the Y-, J-, H,- and K-bands\! \footnote{\url{https://www.eso.org/sci/facilities/paranal/instruments/sphere/inst/filters.html}} of SPHERE-IRDIS. We discuss the results in the next section.

\section{Results and Discussions} \label{sec:rd}

We have presented new models for the scattering polarization detectable from a self-luminous directly imaged Exoplanet arising due to asymmetry caused by its rotation-induced oblateness. We introduced a new formalism to account for the effect of the anisotropic distribution of scatterer in the atmosphere of an oblate planet. It is shown that the emergent flux of an oblate substellar mass object does not differ much from that of a spherical object.  In fact, the flux can be calculated with an insignificant error by assuming the object to be spherical. On the other hand, polarization, being a measure of the anisotropy in the radiation field, strongly depends on the distribution of scatterers and on the shape of the visible disk of the object.  This is clearly demonstrated in Figure~\ref{fig:tdwarf} which shows that the predicted polarization of a T-dwarf can increase significantly when the visible disk is non-spherical while the flux does not alter much. The T-dwarf has been chosen to focus our calculations on a cloud-free atmosphere in order to highlight only the effect of our new three-dimensional approach. Polarization of a cloud-free object arises by atomic and molecular scattering and the angular distribution of the scattered photons is described by the Rayleigh phase matrix.

In the case of the cloudy self-luminous giant planets, we have adopted the atmospheric models of MS11. However, we have introduced a generalized HGR phase matrix in order to treat the scattering due to the cloud particulates more accurately.  Figure~\ref{fig:phasefunc} demonstrates that the generalized scattering phase matrix described in this paper provides a better representation of the asymmetric nature of the angular distribution of the photons before and after the scattering of condensate cloud particles.  Figure~\ref{fig:pfonintvec} shows that although the total intensity does not alter significantly, the polarized intensity is slightly overestimated if the Rayleigh-dominated symmetric approximation of the HGR phase matrix is used. This in turn alters the disk-integrated polarization as shown in Figure~\ref{fig:dip}. Figure~\ref{fig:dip} also shows the flux and the polarization estimated by using the generalized HGR phase matrix as well as the new algorithm of integrating the polarization over the rotation induced oblate disk. 
 
In Figures~\ref{fig:contour5} and \ref{fig:contour10} we present the color maps of the intensity vectors normalized to that at the disk-center for a self-luminous Jupiter-sized planet with $T_{\rm eff}=1000$ K and $g=30$ ms$^{-2}$ rotating with periods $P_{\rm rot}=5$ h and $P_{\rm rot}=10$ h respectively at different values of $i$. These figures help us to visualize the patterns of the total intensity ($I_{\rm obs}$) and the polarized intensity ($Q_{\rm obs}$) across the planetary disk and to understand how they contribute to the disk-integrated flux and polarization. The variation of the intensities in the $M$-direction exhibits limb darkening and an increase in polarized intensity from the disk center to the limb which agrees with Figure~\ref{fig:pfonintvec}. 

The variation in the polarization at the local points across the disk along the $\Phi$-direction determines the amount of the net non-zero disk-integrated polarization. At an inclination of $i=0^\circ$, both Figures~\ref{fig:contour5} and \ref{fig:contour10} show that $I_{\rm obs}$ is uniform along the $\Phi$-direction and $Q_{\rm obs}$ exhibits an exact cosinusoidal pattern (see Equation~\ref{eq:iobs1}) along the $\Phi$-direction. The amount of positive polarization along the Y-axis is exactly nullified by the amount of negative polarization along the X-axis causing the net disk-averaged polarization to be zero. This occurs because of the perfect symmetry about the disk center when the inclination angle is $i=0^\circ$. However, for a given value of $P_{\rm rot}$, the asymmetry in $\Phi$-direction increases with an increasing value of $i$ as the poles shift from the $M=0$. This effect is more as the rotation speed increases (lower value of $P_{\rm rot}$) and hence the oblateness increases.  Figure~\ref{fig:contour5} for $P_{\rm rot}=5$ h) represents such a case of highly oblate object.  In this case, the asymmetry is maximum at $i=90^\circ$ and hence the cancellation is minimum because the positive polarization (along Y-axis) is more than the negative polarization (along X-axis). This causes the disk-integrated polarization to be maximum at $i=90^\circ$ for a  value of the rotation period. This effect, however, reduces for the case of $P_{\rm rot}=10$ h since the rotation-induced oblateness is significantly small ($f\sim0.05$) at this low rotation rate. Figure~ \ref{fig:contour10} does not show significant variation in the pattern of $I_{\rm obs}$ and $Q_{\rm obs}$ over the disk for different values of $i$.  Although in this case, the oblateness is close to that of Jupiter ($\sim0.064$), detecting the disk-averaged degree of polarization ($\sim0.005\%$ for $i=30^\circ$ and $\sim0.02\%$ for $i=90^\circ$; see Figure~\ref{fig:photpol-inc}) of such slowly rotating directly imaged Exoplanets may be quite challenging due to extremely low signal-to-noise ratio.

Figure~\ref{fig:dip-teff-g} shows the spectra of the flux and the polarization observable from young self-luminous giant planets with different effective temperature and surface gravity. It shows that the polarization significantly reduces with the increase in surface gravity for the same value of $T_{\rm eff}$ and $P_{\rm rot}$ as pointed out by MS11. This is also demonstrated in Figure~\ref{fig:photpol-teffg} where the detectable disk-averaged polarizations at different wavelength bands for planets with  different effective temperature $T_{\rm eff}$ and surface gravity $g$ are presented as functions of spin rotation period $P_{\rm rot}$. On the other hand, Figure~\ref{fig:photpol-inc} presents disk-integrated polarization at different wavelength bands that may be detectable from a self-luminous Jupiter-sized planet with $T_{\rm eff}=1000$ K and $g=30$ ms$^{-2}$. In this figure, the amount of polarization of the planet with different rotation periods and different inclination angles ($i$) of the rotation axis with respect to the observer is presented. Clearly, with an increase in $i$, the polarization increases for a given value of $P_{\rm rot}$. The polarization predicted is maximum for $i=90^\circ$ i.e. for an equatorial view when the disk asymmetry is maximum and close to zero for $i=0^\circ$ i.e. for a polar view (see Figure~\ref{fig:geom}) when disk asymmetry is almost absent.  

Figures~\ref{fig:betapic} and \ref{fig:roxs} demonstrate that the polarization due to the oblateness of the planets $\beta$ Pic b and ROXs 42B b can be constrained with the observational reports of their $v_e\sin i$, viz. $25\pm3$ km/s \citep{snellen14} and $9.5_{-2.3}^{+2.1}$ km/s \citep{bryan18} respectively. For a given value of $v_e\sin i$, a lower value of $i$ implies a higher value of the equatorial rotation speed ($v_e$) which causes the planet to be more oblate resulting in a higher value of the detectable degree of polarization. Conversely, for a given value of $v_e$, a decreasing value of $i$ implies declining asymmetry of the visible planetary disk leading to a lower degree of the polarization as evident from Figure~\ref{fig:photpol-inc}. However, the former effect outweighs the latter, causing an overall drop in the detectable polarization with an increase in $i$. The minimum value of $i$ is determined by the stability limit of the spin rotation speed of a planet for a given observed value of $v_e\sin i$. These lower limits of $i$ for the two planets are found to be $\sim20^\circ$ and $\sim14^\circ$ respectively. Thus we have been able to determine both the upper and lower limits of the detectable disk-integrated polarization of these planets by using the reported values of $v_e\sin i$ from observations.

Clearly, the degree of polarization of the two planets predicted by our present model is much below the 1-$\sigma$ non-detection upper limit, viz. 0.18\% and 0.19\%, reported by \cite{jensen20}. These predictions are subject to the particular atmospheric model and cloud model we adopt for a particular planet. With the change in the atmospheric model, e.g. if we assume chemical inequilibrium \citep{madhusudhan20} or if we choose a different value of the sedimentation parameter $f_{\rm sed}$ for the cloud model, the predicted degree of polarization can alter. So, if future observations confirm the amount of polarization to be in the order of the non-detection upper limits set by \cite{jensen20}, we need to adopt a different atmospheric or cloud model or identify other sources of polarization such as cloud band or surface inhomogeneity. Our new technique is capable of performing such analysis by including the inhomogeneity of the atmospheres which we will explore in our future work. However, the present study only focuses on the new 3D technique of accounting for the anisotropy in the atmosphere of the planet introduced by the departure from sphericity due to spin rotation and on the generalized form of the HGR phase matrix on the detectable polarization.

\section{Conclusion} \label{sec:con}

It is foreseeable that imaging polarimetry and in the future, spectropolarimetry in synergy with the photometric and spectroscopic techniques can open the door to the unexplored regime of exoplanets. At present, polarimetry is the only prospective tool that can convey information about the axial tilt of a directly imaged planet. This technique is also useful in excavating information about the deeper layers of the atmospheres, especially about the condensates clouds that cannot be fully probed with the spectroscopic technique. However, the correct interpretation of the polarimetric observations demands self-consistent and extensive work on the development of a forward model that can describe the atmospheric processes correctly and explain the polarization arising from those processes. 

All of our calculations are centered on the atmospheric models and cloud models (fixing f$_{\rm sed}$ at 2) of \cite{marley11, jensen20}. With the change in the models adopted, the predicted polarization would also change. Consequently, polarimetric observations can be utilized to distinguish these models. In the present study, we have explored only a few cases in order to demonstrate our new approach to account for the effect of the oblateness of the fast-rotating directly imaged planets on their observable polarization. This study also presents our generic methodology to calculate the phase matrices for a more accurate representation of the scattering process that predominantly dictates the polarization observable from a cloudy exoplanet. Our work provides a three-dimensional view of the atmosphere of an oblate planet which allows us to calculate the polarization arising from the asymmetry caused by rotation-induced non-sphericity more accurately. This technique will be further applied in our follow-up work to calculate the polarization owing to the inhomogeneous atmospheres of the substellar objects, for example, due to the presence of banded or patchy clouds.

To bring this paper into proper shape, we have intensively used the high-performance computing facility (Delphinus) of Indian Institute of Astrophysics, Bangalore. We are thankful to the computer division of Indian Institute of Astrophysics for the help and cooperation extended for the present project.


\begin{thebibliography}{}
\expandafter\ifx\csname natexlab\endcsname\relax\def\natexlab#1{#1}\fi
\providecommand{\url}[1]{\href{#1}{#1}}
\providecommand{\dodoi}[1]{doi:~\href{http://doi.org/#1}{\nolinkurl{#1}}}
\providecommand{\doeprint}[1]{\href{http://ascl.net/#1}{\nolinkurl{http://ascl.net/#1}}}
\providecommand{\doarXiv}[1]{\href{https://arxiv.org/abs/#1}{\nolinkurl{https://arxiv.org/abs/#1}}}

\bibitem[{{Ackerman} \& {Marley}(2001)}]{ackerman01}
{Ackerman}, A.~S., \& {Marley}, M.~S. 2001, \apj, 556, 872,
  \dodoi{10.1086/321540}

\bibitem[{{Barnes} \& {Fortney}(2003)}]{barnes03}
{Barnes}, J.~W., \& {Fortney}, J.~J. 2003, \apj, 588, 545,
  \dodoi{10.1086/373893}

\bibitem[{{Batalha} {et~al.}(2019){Batalha}, {Marley}, {Lewis}, \&
  {Fortney}}]{batalha19}
{Batalha}, N.~E., {Marley}, M.~S., {Lewis}, N.~K., \& {Fortney}, J.~J. 2019,
  \apj, 878, 70, \dodoi{10.3847/1538-4357/ab1b51}

\bibitem[{{Bryan} {et~al.}(2018){Bryan}, {Benneke}, {Knutson}, {Batygin}, \&
  {Bowler}}]{bryan18}
{Bryan}, M.~L., {Benneke}, B., {Knutson}, H.~A., {Batygin}, K., \& {Bowler},
  B.~P. 2018, Nature Astronomy, 2, 138, \dodoi{10.1038/s41550-017-0325-8}

\bibitem[{{Burgasser} {et~al.}(2002){Burgasser}, {Marley}, {Ackerman},
  {Saumon}, {Lodders}, {Dahn}, {Harris}, \& {Kirkpatrick}}]{burgasser02}
{Burgasser}, A.~J., {Marley}, M.~S., {Ackerman}, A.~S., {et~al.} 2002, \apjl,
  571, L151, \dodoi{10.1086/341343}

\bibitem[{{Chakrabarty} \& {Sengupta}(2020)}]{chakrabarty20}
{Chakrabarty}, A., \& {Sengupta}, S. 2020, \apj, 898, 89,
  \dodoi{10.3847/1538-4357/ab9a33}

\bibitem[{{Chakrabarty} \& {Sengupta}(2021)}]{chakrabarty21}
---. 2021, \apj, 917, 83, \dodoi{10.3847/1538-4357/ac0bb7}

\bibitem[{{Chandrasekhar}(1933)}]{chandrasekhar33}
{Chandrasekhar}, S. 1933, \mnras, 93, 539, \dodoi{10.1093/mnras/93.8.539}

\bibitem[{{Chandrasekhar}(1960)}]{chandrasekhar60}
---. 1960, {Radiative transfer}

\bibitem[{{de Kok} {et~al.}(2011){de Kok}, {Stam}, \& {Karalidi}}]{dekok11}
{de Kok}, R.~J., {Stam}, D.~M., \& {Karalidi}, T. 2011, \apj, 741, 59,
  \dodoi{10.1088/0004-637X/741/1/59}

\bibitem[{{Espinosa Lara} \& {Rieutord}(2011)}]{lara11}
{Espinosa Lara}, F., \& {Rieutord}, M. 2011, \aap, 533, A43,
  \dodoi{10.1051/0004-6361/201117252}

\bibitem[{{Freedman} {et~al.}(2014){Freedman}, {Lustig-Yaeger}, {Fortney},
  {Lupu}, {Marley}, \& {Lodders}}]{freedman14}
{Freedman}, R.~S., {Lustig-Yaeger}, J., {Fortney}, J.~J., {et~al.} 2014, \apjs,
  214, 25, \dodoi{10.1088/0067-0049/214/2/25}

\bibitem[{{Freedman} {et~al.}(2008){Freedman}, {Marley}, \&
  {Lodders}}]{freedman08}
{Freedman}, R.~S., {Marley}, M.~S., \& {Lodders}, K. 2008, \apjs, 174, 504,
  \dodoi{10.1086/521793}

\bibitem[{{James}(1964)}]{james64}
{James}, R.~A. 1964, \apj, 140, 552, \dodoi{10.1086/147949}

\bibitem[{{Jensen-Clem} {et~al.}(2020){Jensen-Clem}, {Millar-Blanchaer}, {van
  Holstein}, {Mawet}, {Graham}, {Sengupta}, {Marley}, {Snik}, {Vigan},
  {Hinkley}, {de Boer}, {Girard}, {De Rosa}, {Bowler}, {Wiktorowicz}, {Perrin},
  {Crepp}, \& {Macintosh}}]{jensen20}
{Jensen-Clem}, R., {Millar-Blanchaer}, M.~A., {van Holstein}, R.~G., {et~al.}
  2020, \aj, 160, 286, \dodoi{10.3847/1538-3881/abc33d}

\bibitem[{{Joseph} {et~al.}(1976){Joseph}, {Wiscombe}, \& {Weinman}}]{joseph76}
{Joseph}, J.~H., {Wiscombe}, W.~J., \& {Weinman}, J.~A. 1976, Journal of
  Atmospheric Sciences, 33, 2452,
  \dodoi{https://doi.org/10.1175/1520-0469(1976)033<2452:TDEAFR>2.0.CO;2}

\bibitem[{{Liu} \& {Weng}(2006)}]{liu06}
{Liu}, Q., \& {Weng}, F. 2006, \ao, 45, 7475, \dodoi{10.1364/AO.45.007475}

\bibitem[{{Lodders}(2010)}]{lodders10}
{Lodders}, K. 2010, Astrophysics and Space Science Proceedings, 16, 379,
  \dodoi{10.1007/978-3-642-10352-0\_8}

\bibitem[{Lodders(2020)}]{lodders20}
Lodders, K. 2020, Solar Elemental Abundances,  Oxford University Press,
  \dodoi{10.1093/acrefore/9780190647926.013.145}.
\newblock
  \url{https://oxfordre.com/planetaryscience/view/10.1093/acrefore/9780190647926.001.0001/acrefore-9780190647926-e-145}

\bibitem[{{Madhusudhan} {et~al.}(2020){Madhusudhan}, {Nixon}, {Welbanks},
  {Piette}, \& {Booth}}]{madhusudhan20}
{Madhusudhan}, N., {Nixon}, M.~C., {Welbanks}, L., {Piette}, A. A.~A., \&
  {Booth}, R.~A. 2020, \apjl, 891, L7, \dodoi{10.3847/2041-8213/ab7229}

\bibitem[{Marley {et~al.}(2018)Marley, Saumon, Morley, \& Fortney}]{marleyz18}
Marley, M., Saumon, D., Morley, C., \& Fortney, J. 2018, {Sonora 2018:
  Cloud-free, solar composition, solar C/O substellar atmosphere models and
  spectra}, nc\_m+0.0\_co1.0\_v1.0,  Zenodo, \dodoi{10.5281/zenodo.1309035}.
\newblock \url{https://doi.org/10.5281/zenodo.1309035}

\bibitem[{Marley {et~al.}(2021)Marley, Saumon, Morley, Fortney, Visscher,
  Freedman, \& Lupu}]{marleyz21}
Marley, M., Saumon, D., Morley, C., {et~al.} 2021, {Sonora Bobcat: cloud-free,
  substellar atmosphere models, spectra, photometry, evolution, and chemistry},
  Sonora Bobcat,  Zenodo, \dodoi{10.5281/zenodo.5063476}.
\newblock \url{https://doi.org/10.5281/zenodo.5063476}

\bibitem[{{Marley} {et~al.}(2002){Marley}, {Seager}, {Saumon}, {Lodders},
  {Ackerman}, {Freedman}, \& {Fan}}]{marley02}
{Marley}, M.~S., {Seager}, S., {Saumon}, D., {et~al.} 2002, \apj, 568, 335,
  \dodoi{10.1086/338800}

\bibitem[{{Marley} \& {Sengupta}(2011)}]{marley11}
{Marley}, M.~S., \& {Sengupta}, S. 2011, \mnras, 417, 2874,
  \dodoi{10.1111/j.1365-2966.2011.19448.x}

\bibitem[{{Marley} {et~al.}(2021){Marley}, {Saumon}, {Visscher}, {Lupu},
  {Freedman}, {Morley}, {Fortney}, {Seay}, {Smith}, {Teal}, \&
  {Wang}}]{marley21}
{Marley}, M.~S., {Saumon}, D., {Visscher}, C., {et~al.} 2021, arXiv e-prints,
  arXiv:2107.07434.
\newblock \doarXiv{2107.07434}

\bibitem[{{Marois} {et~al.}(2008){Marois}, {Macintosh}, {Barman}, {Zuckerman},
  {Song}, {Patience}, {Lafreni{\`e}re}, \& {Doyon}}]{marois08}
{Marois}, C., {Macintosh}, B., {Barman}, T., {et~al.} 2008, Science, 322, 1348,
  \dodoi{10.1126/science.1166585}

\bibitem[{{Miles-P{\'a}ez} {et~al.}(2015){Miles-P{\'a}ez}, {Zapatero Osorio},
  \& {Pall{\'e}}}]{miles15}
{Miles-P{\'a}ez}, P.~A., {Zapatero Osorio}, M.~R., \& {Pall{\'e}}, E. 2015,
  \aap, 580, L12, \dodoi{10.1051/0004-6361/201424626}

\bibitem[{{Miles-P{\'a}ez} {et~al.}(2019){Miles-P{\'a}ez}, {Zapatero Osorio},
  {Pall{\'e}}, \& {Metchev}}]{miles19}
{Miles-P{\'a}ez}, P.~A., {Zapatero Osorio}, M.~R., {Pall{\'e}}, E., \&
  {Metchev}, S.~A. 2019, \mnras, 484, L38, \dodoi{10.1093/mnrasl/slz001}

\bibitem[{{Millar-Blanchaer} {et~al.}(2020){Millar-Blanchaer}, {Girard},
  {Karalidi}, {Marley}, {van Holstein}, {Sengupta}, {Mawet}, {Kataria}, {Snik},
  {de Boer}, {Jensen-Clem}, {Vigan}, \& {Hinkley}}]{millar20}
{Millar-Blanchaer}, M.~A., {Girard}, J.~H., {Karalidi}, T., {et~al.} 2020,
  \apj, 894, 42, \dodoi{10.3847/1538-4357/ab6ef2}

\bibitem[{{Sanghavi} \& {Shporer}(2018)}]{sanghavi18}
{Sanghavi}, S., \& {Shporer}, A. 2018, \apj, 866, 28,
  \dodoi{10.3847/1538-4357/aadf94}

\bibitem[{{Sanghavi} \& {West}(2019)}]{sanghavi19}
{Sanghavi}, S., \& {West}, R. 2019, \apj, 877, 134,
  \dodoi{10.3847/1538-4357/ab1b4e}

\bibitem[{{Saumon} \& {Marley}(2008)}]{saumon08}
{Saumon}, D., \& {Marley}, M.~S. 2008, \apj, 689, 1327, \dodoi{10.1086/592734}

\bibitem[{{Sengupta} {et~al.}(2020){Sengupta}, {Chakrabarty}, \&
  {Tinetti}}]{sengupta20}
{Sengupta}, S., {Chakrabarty}, A., \& {Tinetti}, G. 2020, \apj, 889, 181,
  \dodoi{10.3847/1538-4357/ab6592}

\bibitem[{{Sengupta} \& {Kwok}(2005)}]{sengupta05}
{Sengupta}, S., \& {Kwok}, S. 2005, \apj, 625, 996, \dodoi{10.1086/429659}

\bibitem[{{Sengupta} \& {Marley}(2009)}]{sengupta09}
{Sengupta}, S., \& {Marley}, M.~S. 2009, \apj, 707, 716,
  \dodoi{10.1088/0004-637X/707/1/716}

\bibitem[{{Sengupta} \& {Marley}(2010)}]{sengupta10}
---. 2010, \apjl, 722, L142, \dodoi{10.1088/2041-8205/722/2/L142}

\bibitem[{{Simmons}(1982)}]{simmons82}
{Simmons}, J.~F.~L. 1982, \mnras, 200, 91, \dodoi{10.1093/mnras/200.1.91}

\bibitem[{{Snellen} {et~al.}(2014){Snellen}, {Brandl}, {de Kok}, {Brogi},
  {Birkby}, \& {Schwarz}}]{snellen14}
{Snellen}, I. A.~G., {Brandl}, B.~R., {de Kok}, R.~J., {et~al.} 2014, \nat,
  509, 63, \dodoi{10.1038/nature13253}

\bibitem[{{Stolker} {et~al.}(2017){Stolker}, {Min}, {Stam}, {Molli{\`e}re},
  {Dominik}, \& {Waters}}]{stolker17}
{Stolker}, T., {Min}, M., {Stam}, D.~M., {et~al.} 2017, \aap, 607, A42,
  \dodoi{10.1051/0004-6361/201730780}

\bibitem[{{van Holstein} {et~al.}(2021){van Holstein}, {Stolker},
  {Jensen-Clem}, {Ginski}, {Milli}, {de Boer}, {Girard}, {Wahhaj}, {Bohn},
  {Millar-Blanchaer}, {Benisty}, {Bonnefoy}, {Chauvin}, {Dominik}, {Hinkley},
  {Keller}, {Keppler}, {Langlois}, {Marino}, {M{\'e}nard}, {Perrot}, {Schmidt},
  {Vigan}, {Zurlo}, \& {Snik}}]{holstein21}
{van Holstein}, R.~G., {Stolker}, T., {Jensen-Clem}, R., {et~al.} 2021, \aap,
  647, A21, \dodoi{10.1051/0004-6361/202039290}

\bibitem[{{Xuan} {et~al.}(2020){Xuan}, {Bryan}, {Knutson}, {Bowler}, {Morley},
  \& {Benneke}}]{xuan20}
{Xuan}, J.~W., {Bryan}, M.~L., {Knutson}, H.~A., {et~al.} 2020, \aj, 159, 97,
  \dodoi{10.3847/1538-3881/ab67c4}

\bibitem[{{Zhou} {et~al.}(2016){Zhou}, {Apai}, {Schneider}, {Marley}, \&
  {Showman}}]{zhou16}
{Zhou}, Y., {Apai}, D., {Schneider}, G.~H., {Marley}, M.~S., \& {Showman},
  A.~P. 2016, \apj, 818, 176, \dodoi{10.3847/0004-637X/818/2/176}

\end{thebibliography}

\end{document}